\documentclass[aps,prx,twocolumn,superscriptaddress,showpacs]{revtex4-1}

\usepackage{amsmath,amssymb,amsfonts,bm,braket,color,epsfig,epstopdf,graphicx,multirow,latexsym,verbatim,ulem,tabularx}
\usepackage[colorlinks,linkcolor=blue,citecolor=blue,urlcolor=blue]{hyperref}

\graphicspath{{Figures/}}

\begin{document}
\title{Topological invariants for interacting topological insulators: I. Efficient numerical evaluation scheme and implementations}
\author{Yuan-Yao He}
\author{Han-Qing Wu}
\affiliation{Department of Physics, Renmin University of China, Beijing 100872, China}
\author{Zi Yang Meng}
\affiliation{Beijing National Laboratory
for Condensed Matter Physics, and Institute of Physics, Chinese
Academy of Sciences, Beijing 100190, China}
\author{Zhong-Yi Lu}
\affiliation{Department of Physics, Renmin University of China, Beijing 100872, China}

\begin{abstract}
The aim of this series of two papers is to discuss topological invariants for interacting topological insulators (TIs). In the first paper (I), we provide a paradigm of efficient numerical evaluation scheme for topological invariants, in which we demystify the procedures and techniques employed in calculating $Z_2$ invariant and spin Chern number via zero-frequency single-particle Green's function in quantum Monte Carlo (QMC) simulations. Here we introduce a {\it periodization} process to overcome the ubiquitous finite-size effect, so that the calculated spin Chern number shows ideally quantized values. We also show that making use of symmetry properties of the underlying systems can greatly reduce the computational effort. To demonstrate the effectiveness of our numerical evaluation scheme, especially the {\it periodization} process, of topological invariants, we apply it on two independent two-dimensional models of interacting topological insulators. In the subsequent paper (II), we apply the scheme developed here to wider classes of models of interacting topological insulators, for which certain limitation of constructing topological invariant via single-particle Green's functions will be presented.
\end{abstract}


\date{\today} \maketitle


\section{Introduction}
\label{sec:Introduction}
Topological insulators (TIs)~\cite{Hasan2010,Qi2011} in free fermion systems can be characterized by topological invariants, such as $Z_2$ invariant and spin Chern number, and the calculation of these topological invariants is straightforward via noninteracting Hamiltonian matrix. However, the characterization of interacting topological insulators via these topological invariants is still not well understood, both conceptually and technically in numerical calculation. Of course, there are already various achievements. In the quantum (anomalous) Hall insulators~\cite{Klitzing1980,Haldane1988} with broken time-reversal symmetry, the characterizing topological invariant is the Thouless-Kohmoto-Nightingale-den Nijs (TKNN) invariant~\cite{Thouless1982,Avron1983}, or first Chern number, which takes integer values. For TIs protected by time-reversal symmetry and charge $U(1)$ symmetry, the $Z_2$ topological index is applied to describe the system~\cite{Kane2005a,Kane2005b,Fu2006time}. With additional $U(1)$ spin rotational symmetry in TIs, the $U(1)_{\text{spin}}\times U(1)_{\text{charge}}\rtimes Z_2^{T}$ symmetry ($Z_2^{T}$ stands for time-reversal symmetry) results in $\mathbb{Z}$ classification and the appropriate topological invariant is the spin Chern number $C_s$~\cite{Sheng2006spin,Prodan2009}, which is actually the case for Kane-Mele model without Rashba spin-orbit coupling~\cite{Kane2005a,Kane2005b}. In such systems, the spin Hall conductivity $\sigma_{xy}^{spin}$ is related to spin Chern number as $\sigma_{xy}^{spin}=C_s\frac{e}{2\pi}$.

For noninteracting TIs, both $Z_2$ invariant~\cite{Kane2005b,Fu2006time} and spin Chern Number~\cite{Sheng2006spin,Prodan2009} can be simply evaluated from Hamiltonian matrix in band basis. For example, if a system has spatial inversion symmetry, the $Z_2$ invariant can be calculated as product of parity eigenvalues of all occupied energy bands at all time-reversal invariant momentum (TRIM) points in Brillouin zone (BZ)~\cite{Fu2007b}. Spin Chern number, on the other hand, can be calculated by simply integrating Berry curvature over the Brillouin zone (BZ).

For interacting TIs, the evaluations of topological invariants become much more difficult and subtle. The proposals include constructing topological invariants from single-particle Green's function~\cite{Avron1983,So1985,Ishikawa,Volovik1988,Wang2010,Gurarie2011,volovik2009universe} or imposing twisted boundary phases to the ground state wavefunction~\cite{Niu1985,Wang2014}. Recently, the constructions of topological invariants from single-particle Green's function, especially the zero-frequency version~\cite{Wang2012c,Wang2013topological}, have been actively investigated. Implementations of topological invariants constructed from zero-frequency single-particle Green's function~\cite{Wang2012c} in many-body numerical techniques have been carried out in various studies. In one-dimensional Su-Schrieffer-Heeger model~\cite{Yoshida2014}, the winding number based on zero-frequency single-particle Green's function is calculated by DMRG method in distinguishing topologically non-trivial and trivial phases. By LDA+Gutzwiller and LDA+DMFT methods, the $Z_2$ invariant has been applied in identifying the correlated TIs SmB$_6$~\cite{Lu2013} and PuB$_6$~\cite{Deng2013Plutonium}. The $Z_2$ invariant has also been calculated in QMC~\cite{Hung2013,Lang2013,Hung2014,Meng2014} and cluster perturbation theory~\cite{Grandi2015a,Grandi2015b} for various generalizations of the Kane-Mele-Hubbard model. Moreover, there are dynamical mean-field theory calculations of Bernevig-Hughes-Zhang model with interactions in which the $Z_2$ invariant is calculated~\cite{Budich2012fluctuation,Budich2013,Amaricci2015}. As for spin Chern number constructed from zero-frequency single-particle Green's function, it has been applied to verify the topological phase transitions in Kane-Mele-Hubbard model by quantum Monte Carlo~\cite{Hung2014,Meng2014} and cellular dynamical mean-field theory (CDMFT)~\cite{Chen2015}.

The issue of evaluating spin Chern number for interacting topological insulators in CDMFT~\cite{Chen2015}, or more generally quantum cluster methods~\cite{Maier2005}, is that the correlation effects can only be captured inside the small cluster, even through the calculated spin Chern number is quantized due to the mean-field bath at thermodynamic limit. Nevertheless, such approaches cannot faithfully monitor the topological phase transitions where a length scale larger than the cluster size is involved. On the other hand, QMC is more accurate in capturing both short- and long-range correlation effects by handling supercell with much larger size, but, in obtaining topological invariants, it suffers from finite-size effect and the topological invariants calculated from QMC for interacting TIs are not exactly integer-quantized~\cite{Hung2013,Hung2014}. Actually, due to finite size effect, spin Chern number can be even far away from expected integer results~\cite{Hung2014}. Thus, it will be a great improvement if one can bring the merit of CDMFT (its thermodynamic limit) into QMC to overcome the finite-size effect in topological invariants, since the integer quantization of topological invariants is essential for achieving well-defined topological phases and identifying topological phase transitions. Here, we provide such a scheme.


In this work, employing large-scale quantum Monte Carlo simulations for interacting TIs, we eliminate the severe finite-size effect and obtain quantized topological invariants by introducing a {\it periodization} scheme and imposing symmetries of the studied systems. The numerical evaluation scheme of both $Z_2$ invariant and spin Chern number constructed from zero-frequency single-particle Green's function proposed in Ref.~\cite{Wang2012c} is systematically presented with all important details. To demonstrate the strength of our calculation scheme, especially the {\it periodization} process, we test it on two independent 2D models of interacting TIs, in which the topological phase transitions driven by one-body model parameters are detected by the integer-quantized topological invariants.

The rest of the paper is organized as follows. In Sec.~\ref{sec:TopoInvQMCMethod}, despite of the already existing literature~\cite{Wang2010,Gurarie2011,Wang2011,Wang2012a,Wang2012b,Wang2012c,Wang2014}, the construction of the $Z_2$ invariants and spin Chern number via zero-frequency single-particle Green's function is discussed for the sake of a self-contained narrative. A brief introduction of projector Quantum Monte Carlo method is also presented. Then our numerical calculation scheme and the {\it periodization} technique for these topological invariants are introduced in detail in Sec.~\ref{sec:NumEva}. After that, we show the applications of our scheme in Sec.~\ref{sec:ApplyQMC}, based on QMC simulations for two independent 2D model systems of interacting TIs. Finally, Sec.~\ref{sec:Conclusion} summarizes our method and makes connection with the paper (II) in this series on identifying interaction-driven topological phase transitions without symmetry breaking by the topological invariants calculated via the scheme presented here for interacting topological insulators, where the limitation of topological invariants constructed from the single-particle Green's function is clearly manifested.

\section{Topological Invariants and Quantum Monte Carlo method}
\label{sec:TopoInvQMCMethod}

\subsection{$Z_2$ invariant and spin Chern number}
\label{sec:TopoInv}

In interacting fermion systems, the single-particle Green's function is given as $G(i\omega,\mathbf{k})=[i\omega+\mu-\mathcal{H}(\mathbf{k})-\Sigma(i\omega,\mathbf{k})]^{-1}$, where $\Sigma(i\omega,\mathbf{k})$ is the self-energy originating
from interaction. The zero-frequency single-particle Green's function is Hermitian~\cite{Wang2012c} and we can obtain its real eigenvalues by diagonalizing the Green's function matrix $G(i\omega=0,\mathbf{k})_{N_{O}\times N_{O}}$, where $N_{O}$ is the number of orbitals or bands. In the language of Ref.~\cite{Wang2013topological}, the topological invariants can be defined as follows. We simply define the so-called topological Hamiltonian $h_\textrm{t}(\mathbf{k}) = -G^{-1}(i\omega=0,\mathbf{k})_{N_{O}\times N_{O}}$, and then calculate the topological invariant as if $h_\textrm{t}(\mathbf{k})$ is a
noninteracting Bloch Hamiltonian. The advantage of $h_\textrm{t}(\mathbf{k})$ is that it reduces to the free Bloch Hamiltonian in the non-interacting limit. Equivalently, we can work with $G(i\omega=0,\mathbf{k})_{N_{O}\times N_{O}}$, since the eigenvectors of $G(i\omega=0,\mathbf{k})_{N_{O}\times N_{O}}$ and $-G^{-1}(i\omega=0,\mathbf{k})_{N_{O}\times N_{O}}$ are the same.

In this paper we study systems with both time-reversal symmetry and $U(1)$ spin-rotational symmetry, indicating $S_z$ conservation. Taking the $U(1)$ spin rotational symmetry into account, we can see that the Green's function is diagonal with respect to the spin index, and the two sub-matrices are denoted as $G_\sigma$ ($\sigma=\uparrow,\downarrow$), each of which can be diagonalized as $G_{\sigma}(0,\mathbf{k})|\phi_m(0,\mathbf{k})\rangle=\mu_m(0,\mathbf{k})|\phi_m(0,\mathbf{k})\rangle$. Both $Z_2$ invariant and spin Chern number can be simply constructed from the eigenvectors $|\phi_m(0,\mathbf{k})\rangle$. For time-reversal invariant and spatial-inversion-symmetric systems with interactions, the $Z_2$ invariant can be constructed from $|\phi_m(0,\mathbf{k})\rangle$ at TRIM points. The $Z_2$ invariant can be expressed as~\cite{Wang2012c}
\begin{eqnarray}
\label{eq:Z2Int}
(-1)^\nu=\prod_{\boldsymbol{\kappa}\in \text{TRIM}}\prod_{\mu_m>0} \eta_{m}(\boldsymbol{\kappa}),
\end{eqnarray}
with $\eta_{m}(\boldsymbol{\kappa})=\langle\phi_m(0,\boldsymbol{\kappa}) | \hat{P} |\phi_m(0,\boldsymbol{\kappa})\rangle$, $\boldsymbol{\kappa}$ stands for TRIM points and $\hat{P}$ is the spatial inversion symmetry operator. Here we have already taken the Kramer's degeneracy at TRIM points into consideration, and we only incorporates the parity in one spin sector in Eq.~(\ref{eq:Z2Int}). This expression of $Z_2$ invariant is a generalization of that for free fermion system to interacting systems. Note that $\{|\phi_m \rangle\}$ ($\mu_m>0$) reduces to the filled Bloch bands in the noninteracting limit. Numerical evaluation of this $Z_2$ invariant in correlated systems is quite simple, and it has been demonstrated~\cite{Budich2013,Meng2014,Hung2013,Lang2013,Hung2014,Chen2015} that this topological invariant works well in detecting topological phase transition with change of $Z_2$ invariant in weakly correlated systems.

So far we only discussed the $Z_2$ invariant. Due to the $U(1)$ spin-rotational symmetry, there is actually a $Z$ invariant, which contains more information. To introduce this $Z$ invariant, let us recall the TKNN or Chern number of fermion systems with broken time-reversal symmetry, which has been generalized to interacting fermion systems as~\cite{Wang2012c}
\begin{eqnarray}
\label{eq:ChernIntr1} \centering
\mathcal{C}=\frac{1}{2\pi}\iint_{\mathbf{k}\in
BZ}d^2\mathbf{k}\mathcal{F}_{xy}(\mathbf{k})
\end{eqnarray}
with
\begin{eqnarray}
\label{eq:ChernIntr2}
\centering
&& \mathcal{F}_{xy}(\mathbf{k})=\partial_{k_x}\mathcal{A}_y(\mathbf{k})-\partial_{k_y}\mathcal{A}_x(\mathbf{k}) \nonumber \\
&& \mathcal{A}_i(\mathbf{k})=-i\sum_{\mu_m>0}\langle \phi_m(0,\mathbf{k})|\partial_{k_i}|\phi_m(0,\mathbf{k})\rangle .
\end{eqnarray}
Due to the $U(1)$ spin rotational symmetry, the Green's function is diagonal with respect to the spin index, thus we can calculate Chern numbers by Eq.~(\ref{eq:ChernProject}) for both spin-up and spin-down sectors denoted as $C_{\uparrow}$ and $C_{\downarrow}$. Then the spin Chern number $C_s$ is simply defined as $C_s=(C_{\uparrow}-C_{\downarrow})/2$. For 2D time-reversal invariant systems, $C_{\uparrow}+C_{\downarrow}=0$, which results in the relation $C_s=C_{\uparrow}$, and we only need to calculate Chern number $C_{\uparrow}$ from the spin-up part of zero-frequency single-particle Green's function $G_{\uparrow}(0,\mathbf{k})$. This
expression of spin Chern number has also been applied in QMC~\cite{Hung2014} and CDMFT~\cite{Chen2015} simulations for interacting two-dimensional topological insulators (or the ``quantum spin Hall insulators (QSHI)'', in the older terminology).

Through a simple derivation from Eq.~(\ref{eq:ChernIntr1}) and Eq.~(\ref{eq:ChernIntr2}), we can arrive at an expression that is numerically more convenient~\cite{Avron1983,Hung2014}:
\begin{eqnarray}
\label{eq:ChernProject} \mathcal{C}=&&\frac{1}{2\pi
i}\iint_{\mathbf{k}\in BZ}dk_xdk_y\cdot  \\ \nonumber
&&\textrm{Tr}\Big\{P(\mathbf{k})\left[\partial_{k_x}P(\mathbf{k})\partial_{k_y}P(\mathbf{k})-\partial_{k_y}P(\mathbf{k})\partial_{k_x}P(\mathbf{k})\right]\Big\},
\end{eqnarray}
where $P(\mathbf{k})$ is a projection operator matrix constructed from eigenvectors $|\phi_m(0,\mathbf{k})\rangle$ of $G_{\sigma}(0,\mathbf{k})$:
\begin{eqnarray}
\label{eq:ProjectOperator}
P(\mathbf{k})=\sum_{\mu_m>0}|\phi_m(0,\mathbf{k})\rangle \langle\phi_m(0,\mathbf{k})|.
\end{eqnarray}
The systems we are dealing with in this paper are all based on honeycomb lattice, the detailed implementation of Eq.~(\ref{eq:ChernProject}) on the honeycomb lattice geometry is presented in Appendix~\ref{sec:appendix_a}.

Turning off interactions, we can observe that Eq.~(\ref{eq:Z2Int}) reduces to the $Z_2$ invariant for free fermion systems defined by Fu and Kane~\cite{Fu2007b}, and both Eq.~(\ref{eq:ChernIntr1}) and Eq.~(\ref{eq:ChernProject}) reduce to the TKNN invariant (or Chern number)~\cite{Thouless1982,Avron1983} for noninteracting systems.

In the next section, we introduce our numerical evaluation scheme of $Z_2$ invariant in Eq.~(\ref{eq:Z2Int}) and spin Chern number in Eq.~(\ref{eq:ChernProject}) in QMC simulations, for model systems of interacting TIs with $U(1)_{spin}\times U(1)_{charge}\rtimes Z_2^{T}$ symmetry and the spatial inversion symmetry. Especially, both $Z_2$ invariant and spin Chern number for interacting TIs are necessary to be quantized to achieve well-defined topological phases. Thus, both {\it periodization} technique and implementation of symmetry properties during numerical calculations to reach quantized topological invariants are mainly introduced in next section.

\subsection{Quantum Monte Carlo method}
\label{sec:QMCMethod}

In this series of work, we apply the projector quantum Monte Carlo (PQMC) simulation~\cite{Meng2014}, which is the zero-temperature version of determinantal QMC algorithm~\cite{AssaadEvertz2008}. PQMC method obtains the ground-state observables by carrying out an imaginary time evolution starting from trial wavefunction that has overlap to the true many-body ground state. The ground-state expectation value of physical observable is calculated as follows,
\begin{equation}
\langle \hat{O}\rangle =\lim\limits_{\Theta\to +\infty}
\frac{\langle
\psi_T|e^{-\Theta\hat{H}/2}\hat{O}e^{-\Theta\hat{H}/2}|\psi_T\rangle}{\langle
\psi_T|e^{-\Theta \hat{H}}|\psi_T\rangle},
\label{eq:PQMC_Observable}
\end{equation}
where $|\psi_T\rangle$ is the trial wave function and $\Theta$ is projection parameter. In all the simulations, to ensure that the algorithm arrives at the truly converged ground state of finite size systems, we choose $\Theta=40/t$ and $\Delta\tau=0.05/t$, in which $\Delta\tau$ is the finite imaginary time step applied in the Trotter decomposition of partition function. 

We can obtain both the static and dynamic observables. Static ones include the expectation values of energy densities, double occupancy, and spin-spin correlation function. As for the dynamic properties, we can measure the dynamic single-particle Green's function and spin-spin correlation function, from which we can determine both the single-particle gap and spin gap for the many-body systems. Especially, we concentrate on single-particle Green's function $G_{\sigma}(\mathbf{k},\tau)$ to calculate the topological invariants. Generally, $G_{\sigma}(\mathbf{k},\tau)$ (in spin sector $\sigma$) is defined as
\begin{equation}
[G_{\sigma}(\tau,\mathbf{k})]_{pq} =\frac{1}{N}\sum_{i,j}e^{i\mathbf{k}\cdot(\mathbf{R}_i-\mathbf{R}_j)}\langle
T_{\tau}[c_{ip,\sigma}(\tau)c^{\dagger}_{jq,\sigma}]\rangle,
\label{eq:PQMC_GreenFucTau}
\end{equation}
where $i,j \in[1,N]$ are the unit cell indices and $p,q \in[1,N_{o}]$ are the orbital indices inside a unit cell. In this manner, for each $\mathbf{k}$ point, $G_{\sigma}(\tau,\mathbf{k})$ is a $N_o\times N_o$ hermitian matrix (according to Eq.~\ref{eq:Relation1}), and for the $L\times L$ system, there are $N=L^2$ momentum points. From $G_{\sigma}(\tau,\mathbf{k})$ data, we can directly obtain the single-particle gap. Most importantly, we need to construct the zero-frequency single-particle Green's function $G_{\sigma}(i\omega=0,\mathbf{k})$ from $G_{\sigma}(\tau,\mathbf{k})$ by combining Fourier transformation and symmetry analysis presented in Sec.~\ref{sec:NumEva}, after which the topological invariants can be calculated according to Eq.~\ref{eq:Z2Int} and Eq.~\ref{eq:ChernProject}.

\section{Numerical evaluation scheme of topological invariants}
\label{sec:NumEva}

This section is divided into four successive parts. First, in Sec.~\ref{sec:GIw0k}, we comment on the condition for zero-frequency single-particle Green's function $G_{\sigma}(i\omega=0,\mathbf{k})$ to be well-behaved in both free and interacting fermion systems. Second, in Sec.~\ref{sec:CalculateGIw0k}, we explain how to obtain correct $G_{\sigma}(i\omega=0,\mathbf{k})$ data from QMC simulation on a finite-size system. Third, Sec.~\ref{sec:Numerical} clarifies some numerical details in evaluations of $Z_2$ invariant and spin Chern number. Finally, as the most important part, the {\it periodization} scheme of $G_{\sigma}(i\omega=0,\mathbf{k})$ is introduced in Sec.~\ref{sec:Periodization} to achieve the ideally quantized topological invariants from QMC simulation in a finite-size system.

\subsection{Condition for well-behaved $G(i\omega=0,\mathbf{k})$}
\label{sec:GIw0k}

In this work, we shall be concerned with quantum many-body systems at zero temperature. Suppose that the system under consideration is gapped and has a unique ground state under periodic boundary condition (no intrinsic topological order), thus there is a many-body energy gap $\Delta$ between the ground state energy level $E_0$ and the first excited energy level $E_1$, namely, $\Delta=E_1-E_0$. In the Lehmann representation, we can see that the retarded Green's function $G(z,\mathbf{k})$ with complex frequency variable $z=\omega_R+i\omega_I$ is an analytical function in the $\omega_R\in(-\Delta,\Delta)$ region. In fact, the Lehmann representation for $G(z,\mathbf{k})$ under zero temperature reads
\begin{eqnarray}
\label{eq:GiwkLehmann1} && G_{\alpha\beta}(z,\mathbf{k}) \nonumber \\
 && =\sum_{m\ne 0} \left[\frac{\langle 0|c_{\mathbf{k}\alpha}
|m\rangle \langle m|c_{\mathbf{k}\beta}^{\dagger}
|0\rangle}{z-(E_m-E_0)} + \frac{\langle m|c_{\mathbf{k}\alpha} |0\rangle
\langle 0|c_{\mathbf{k}\beta}^{\dagger} |m\rangle}{z+(E_m-E_0)} \right]. \hspace{0.3cm}
\end{eqnarray}
Since the condition $E_m-E_0\ge\Delta$ is satisfied, $G(z,\mathbf{k})$ has no poles in the $\omega_R\in(-\Delta,\Delta)$ region on the real axis. We can further observe that $G(z,\mathbf{k})$ is actually an analytical function with $\omega_R\in(-\Delta,\Delta)$ and arbitrary $\omega_I$. Topological invariants are defined in terms of zero-frequency Green's function, namely, the Green's function at $z=0$. More precisely, since the chemical potential has been absorbed into the definition of $E_m$ (namely, $E_m$ is the eigenvalue of $\hat{H}-\mu\hat{N}$, where $\hat{N}$ is the particle number operator and $\mu$ is the chemical potential), the zero-frequency refers to the energy exactly at the chemical potential~\cite{Wang2012c}. Tuning chemical potential within the energy gap is harmless in an insulator. Thus, the zero-frequency single-particle Green's function is well-defined for TIs with the above mentioned properties under zero-temperature and one can construct topological invariants from it.

From both Eq.~(\ref{eq:Z2Int}) and Eq.~(\ref{eq:ChernProject}), we can determine that there are two scenarios of the jumping of topological invariants determined in terms of Green's function. The first scenario is the pole of Green's function, which is the conventional case, and the topological transitions in noninteracting fermion systems belong to this class. The second scenario is the zero of Green's function (namely, an eigenvalue of the Green's function matrix becomes zero). From the Lehmann representation, we can see that the first scenario must describe a phase transition, since $E_m-E_0=0$ implies gap closing. On the other hand, the second scenario can be a topological phase transition~\cite{Gurarie2011,Yoshida2014}, but it is not necessarily so.


\subsection{Calculations of $G(i\omega=0,\mathbf{k})$}
\label{sec:CalculateGIw0k}

In QMC, we measure the imaginary-time displaced Green's function $G(\tau,\mathbf{k})$ and obtain $G(i\omega_n,\mathbf{k})$ by Fourier transformation as follows
\begin{eqnarray}
\label{eq:OrignFourier} G(i\omega_n,\mathbf{k})=\int_{0}^{\beta}d\tau
e^{i\omega_n\tau}G(\tau,\mathbf{k}).
\end{eqnarray}
Note that Eq.~(\ref{eq:OrignFourier}) has already incorporated the anti-periodic condition for $G(\tau,\mathbf{k})$ as $G(\tau+\beta,\mathbf{k})=-G(\tau,\mathbf{k})$. However, Eq.~(\ref{eq:OrignFourier}) is only valid for finite temperature. At zero-temperature, $i\omega_n$ becomes continuous on the imaginary frequency axis and the anti-periodic condition for $G(\tau,\mathbf{k})$ is not quite meaningful. In such case, one needs to apply the following Fourier transformation
\begin{eqnarray}
\label{eq:NewFourier1}
G(i\omega,\mathbf{k})=\int_{-\infty}^{+\infty}d\tau e^{i\omega\tau}G(\tau,\mathbf{k}).
\end{eqnarray}
The validity of using Eq.~(\ref{eq:NewFourier1}) under zero-temperature is presented in Appendix~\ref{sec:appendix_b}. From Eq.~(\ref{eq:NewFourier1}), we can directly obtain the zero-frequency single-particle Green's function $G(i\omega,\mathbf{k})$ by substituting $i\omega=0$ as,
\begin{eqnarray}
\label{eq:NewFourier2}
G(i\omega=0,\mathbf{k})=\int_{-\infty}^{+\infty}d\tau G(\tau,\mathbf{k}).
\end{eqnarray}
One can furthermore make use of the symmetry properties of $G(\tau,\mathbf{k})$ to simplify the calculation. For a general multi-band systems with $N_{O}$ orbitals (in each spin sector within a unit cell), one can prove
\begin{eqnarray}
\label{eq:Relation1}
[G_{\sigma}(\tau,\mathbf{k})]_{pq}=[G_{\sigma}(\tau,\mathbf{k})]_{qp}^{\star}
\end{eqnarray}
with $p,q\in[1,N_{O}]$. Eq.~(\ref{eq:Relation1}) explicitly shows that $G_{\sigma}(\tau,\mathbf{k})$ is a $N_{O}\times N_{O}$ Hermitian matrix. If the system preserves spatial inversion symmetry and the corresponding spatial inversion operation transforms $p$ sublattice to $p^\prime$ sublattice, then we can prove
\begin{eqnarray}
\label{eq:Relation2}
[G_{\sigma}(\tau,\mathbf{k})]_{pq}=[G_{\sigma}(\tau,-\mathbf{k})]_{p^\prime q^\prime}.
\end{eqnarray}
Eq.~(\ref{eq:Relation2}) explicitly connects the $G_{\sigma}(\tau,\mathbf{k})$ data with opposite wavevector points. If the system preserves particle-hole symmetry, we can prove
\begin{eqnarray}
\label{eq:Relation3}
[G_{\sigma}(\tau,\mathbf{k})]_{pq}=-\xi_p\xi_q[G_{\sigma}(-\tau,-\mathbf{k})]_{qp},
\end{eqnarray}
where $\xi_p$ and $\xi_q$ are the sign change during on-site particle-hole transformation
as $c_{p}\to\xi_pc_{p}^{\dagger}$ and $c_{q}\to\xi_qc_{q}^{\dagger}$. Eq.~(\ref{eq:Relation3})
shows the connections of $G_{\sigma}(\tau,\mathbf{k})$ data with positive and negative $\tau$.
The detailed proof for Eq.~(\ref{eq:Relation1}), Eq.~(\ref{eq:Relation2}) and Eq.~(\ref{eq:Relation3})
is demonstrated in Appendix~\ref{sec:appendix_c}. Combining these three symmetry properties of the
$G_{\sigma}(\tau,\mathbf{k})$ matrix, we can determine that the number of non-zero and independent
matrix elements in $G_{\sigma}(\tau,\mathbf{k})$ is much smaller than $N_{O}^2$.

\subsection{Numerical details in evaluating topological invariants}
\label{sec:Numerical}

Besides the discussions in Sec.~\ref{sec:GIw0k} and Sec.~\ref{sec:CalculateGIw0k}, there are still some important details in numerical application of Eq.~(\ref{eq:ChernProject}) for calculating the topological invariants.

Firstly, the infinite integral over imaginary time $\tau$ in Eq.~(\ref{eq:NewFourier2}) can be approximated by a cutoff $\theta$ as
\begin{eqnarray}
\label{eq:NewFourier3}
G(i\omega=0,\mathbf{k})\approx\int_{-\theta}^{+\theta}d\tau G(\tau,\mathbf{k}).
\end{eqnarray}
For TIs, the systems have $G_{\sigma}(\tau,\mathbf{k})\propto e^{-\Delta_{sp}(\mathbf{k})\tau}$ at large $\tau$ with $\Delta_{sp}(\mathbf{k})$ as the single-particle gap at $\mathbf{k}$ point. If the gap $\Delta_{sp}(\mathbf{k})$ is large, the exponential decay of $G_{\sigma}(\tau,\mathbf{k})$ in imaginary time will be very fast, and a finite $\theta$ is sufficient for the system to evolve below the energy scale of $\Delta_{sp}$. However, this approximation can induce considerable error around the topological phase transition with single-particle gap closing, since the decaying of $G_{\sigma}(\tau,\mathbf{k})$ is very slow around the transition point. Later on, one will observe the nonmonotonic behavior in results of Chern number $C_{\uparrow}$ close to the topological phase transition points in Sec.~\ref{sec:ApplyQMC}, which is originating from the above $\tau$ cutoff. We then calculate the integral in Eq.~(\ref{eq:NewFourier3}) numerically by a simple trapezoidal method, and the step size for the this simple method is actually the $\Delta\tau$ used in QMC simulations.

Secondly, the first-order derivatives over $k_x, k_y$ in Eq.~(\ref{eq:ChernProject}) are replaced by first-order finite-difference~\cite{Hung2014}. Since the interval between two adjacent $\mathbf{k}$ points is proportional to $1/L$ for a $L\times L$ system, replacing the derivatives by finite-difference will bring in error proportional to $1/L$, which is the key origin for finite-size effect in Chern number calculated by Eq.~(\ref{eq:ChernProject}). Also, the integral over $\mathbf{k}$ in BZ region in Eq.~(\ref{eq:ChernProject}) can only be performed by summations over $L^2$ discrete $\mathbf{k}$ points for a $L\times L$ system. We can imagine this can also contribute to finite-size effect in Chern number calculation by Eq.~(\ref{eq:ChernProject}). Overall, due to the finite-size effect in QMC simulations, the spin Chern number result can be very far from the expected integer~\cite{Hung2014}.

\subsection{Periodization of $G(i\omega=0,\mathbf{k})$}
\label{sec:Periodization}

As discussed above, the topological invariants calculated from QMC simulations in finite-size system suffer severe finite-size effect, and the obtained result is usually far away from the expected integer values. To remove the finite-size effect, we combine the idea of {\it periodization} process as the last step in CDMFT algorithm with QMC simulations. CDMFT simulates a correlated system within a cluster (with some other noninteracting bath energy levels) and open boundary condition is applied for the finite-size cluster. To recover the translational symmetry broken by the open boundary condition, a periodization process~\cite{Parcollet2004,Biroli2004,Sakai2012,QingXiao2015} is applied to construct the Green's function and self-energy with arbitrary momentum resolution. The periodization process is generally realized by
\begin{eqnarray}
\label{eq:PeriodGIw0k0}
Q^L(i\omega_n,\mathbf{k})=\frac{1}{N_c}\sum_{i,j=1}^{N_c}Q_{ij}^C(i\omega_n)e^{i\mathbf{k}\cdot(\mathbf{r}_i-\mathbf{r}_j)},
\end{eqnarray}
where $N_c$ is the size of cluster, $Q^C(i\omega_n)$ is the quantity calculated on the cluster and $Q^L(\mathbf{k},i\omega_n)$ is the periodization result with arbitrary $\mathbf{k}$ point in BZ. In CDMFT, generally one can choose $Q^C(i\omega_n)$ to be $M^{C}(i\omega_n)$ and $\Sigma^{C}(i\omega_n)$, where $M^{C}(i\omega_n)=[i\omega_n+\mu-\Sigma^{C}(i\omega_n)]^{-1}$ is the cluster cumulant and $\Sigma^{C}(i\omega_n)$ is the cluster self-energy. Naively thinking, directly substituting $G^{C}(i\omega_n)$ into Eq.~(\ref{eq:PeriodGIw0k0}) seems to be the simplest way, since matrix inverse operations are needed during numerical calculations if $Q^C(i\omega_n)$ is chosen to be $M^{C}(i\omega_n)$ and $\Sigma^{C}(i\omega_n)$ instead. However, due to the breaking of lattice translational symmetry, the simple $G$ periodization can generate artificial results and another more complicated version of $G$ periodization instead of Eq.~(\ref{eq:PeriodGIw0k0}) should be applied to obtain reasonable results~\cite{Sakai2012}. For our purpose of calculating topological invariants in QMC, nevertheless, we can apply the $G$ periodization method simply via Eq.~(\ref{eq:PeriodGIw0k0}) by choosing $Q^C(i\omega_n)$ to be $G^{C}(i\omega_n)$, since the QMC simulations respect periodic boundary condition on finite-size system, which serves as the difference between the periodization methods applied here and CDMFT~\cite{Sakai2012}.

The {\it periodization} process is performed as follows. We first carry out QMC simulation for $L\times L$ system with $N_{O}$ orbitals inside a unit cell (for one spin sector). Then with Eq.~(\ref{eq:NewFourier3}) we obtain the $G_{\sigma}(i\omega=0,\mathbf{k})$ data at $L^2$ discrete momentum points in BZ. After that we carry out the Fourier transformation to get the real space $[G_{\sigma,ij}(i\omega=0)]_{pq}$ data. Finally, we apply the {\it periodization} process as Eq.~(\ref{eq:PeriodGIw0k0}) to obtain $G_{\sigma}(i\omega=0,\mathbf{k})$ data at arbitrary $\mathbf{k}$ point in BZ as
\begin{eqnarray}
\label{eq:PeriodGIw0k1}
[G_{\sigma}(0,\mathbf{k})]_{pq}=\frac{1}{L^2}\sum_{i,j=1}^{L^2}[G_{\sigma,ij}(i\omega=0)]_{pq}e^{i\mathbf{k}\cdot(\mathbf{R}_i-\mathbf{R}_j)}, \hspace{0.5cm}
\end{eqnarray}
where $i,j$ stands for unit cells and $p,q\in[1,N_{O}]$ are the indexes for orbital or band. Throughout the {\it periodization} process, the
$G_{\sigma}(i\omega=0,\mathbf{k})$ data at the original $L^2$ wavevector points keep unchanged. Based on this fact, we can take the {\it periodization} process in Eq.~(\ref{eq:PeriodGIw0k1}) as Lagrange's interpolation of $G_{\sigma}(i\omega=0,\mathbf{k})$. The validity of the {\it periodization} process is based on the observation that the topology of the system will not change during a continuous deformation if there is no gap closing, which makes the {\it periodization} process natural and appropriate. This means that if the continuous deformation do not generate any singularity of the single-particle Green's function, the {\it periodization} process will not alter the topological property of the system, but only improve the quality of the calculated topological invariants and allow us to achieve quantized topological invariants, especially the spin Chern number.


During numerical calculation, the lattice size for {\it periodization} (interpolation) process will be large but still discrete, so we choose the interpolation lattice size $IL\times IL$ and construct the $G_{\sigma}(i\omega=0,\mathbf{k})$ data at $IL\times IL$ wavevector points in BZ by Eq.~(\ref{eq:PeriodGIw0k1}). $IL$ can be much bigger than original lattice size $L$ in QMC simulation. Then we evaluate the spin Chern number for the $IL\times IL$ system by Eq.~(\ref{eq:ChernProject}). In next section, one can clearly observe the converging of spin Chern number to the ideal quantized value with increasing $IL$.

\section{Applications in QMC}
\label{sec:ApplyQMC}

In this section, we apply the above mentioned numerical scheme to two independent 2D interacting TIs. The topological phase transitions in these systems are driven by single-particle parameters. 
Results of topological invariants across topological phase transitions in generalized Kane-Mele-Hubbard model (GKMH) are shown in Sec.~\ref{sec:GeneKMH}, while Sec.~\ref{sec:ClusterKMH} concentrates on topological phase transitions in cluster Kane-Mele-Hubbard model (CKMH).

We set the following parameters in the QMC simulations, $\Theta=40/t,\Delta\tau=0.05/t$. For the imaginary time integration in Eq.~(\ref{eq:NewFourier3}), a cutoff $\theta=20/t$ is applied.

\subsection{Generalized Kane-Mele-Hubbard model}
\label{sec:GeneKMH}
Generalized Kane-Mele-Hubbard (GKMH) model~\cite{Hung2013,Hung2014,Meng2014} is given by
\begin{eqnarray}
\hat{H} &=& -t\sum_{\langle i,j\rangle\sigma}t_{ij}(c^{\dagger}_{i\sigma}c_{j\sigma} + h.c) -t_3\sum_{\langle\!\langle\!\langle i,j\rangle\!\rangle\!\rangle\sigma}(c^{\dagger}_{i\sigma}c_{j\sigma} + h.c.) \nonumber\\
&& +i\lambda\sum_{\langle\!\langle i,j \rangle\!\rangle \alpha\beta}v_{ij}(c^{\dagger}_{i\alpha}\sigma^{z}_{\alpha\beta}c_{j\beta}
-c^{\dagger}_{j\beta}\sigma^{z}_{\beta\alpha}c_{i\alpha}) \nonumber\\
&& +\frac{U}{2}\sum_{i}(n_{i\uparrow}+n_{i\downarrow}-1)^2,
\label{eq:GKMH}
\end{eqnarray}
\begin{figure}[t]
\centering
\includegraphics[width=\columnwidth]{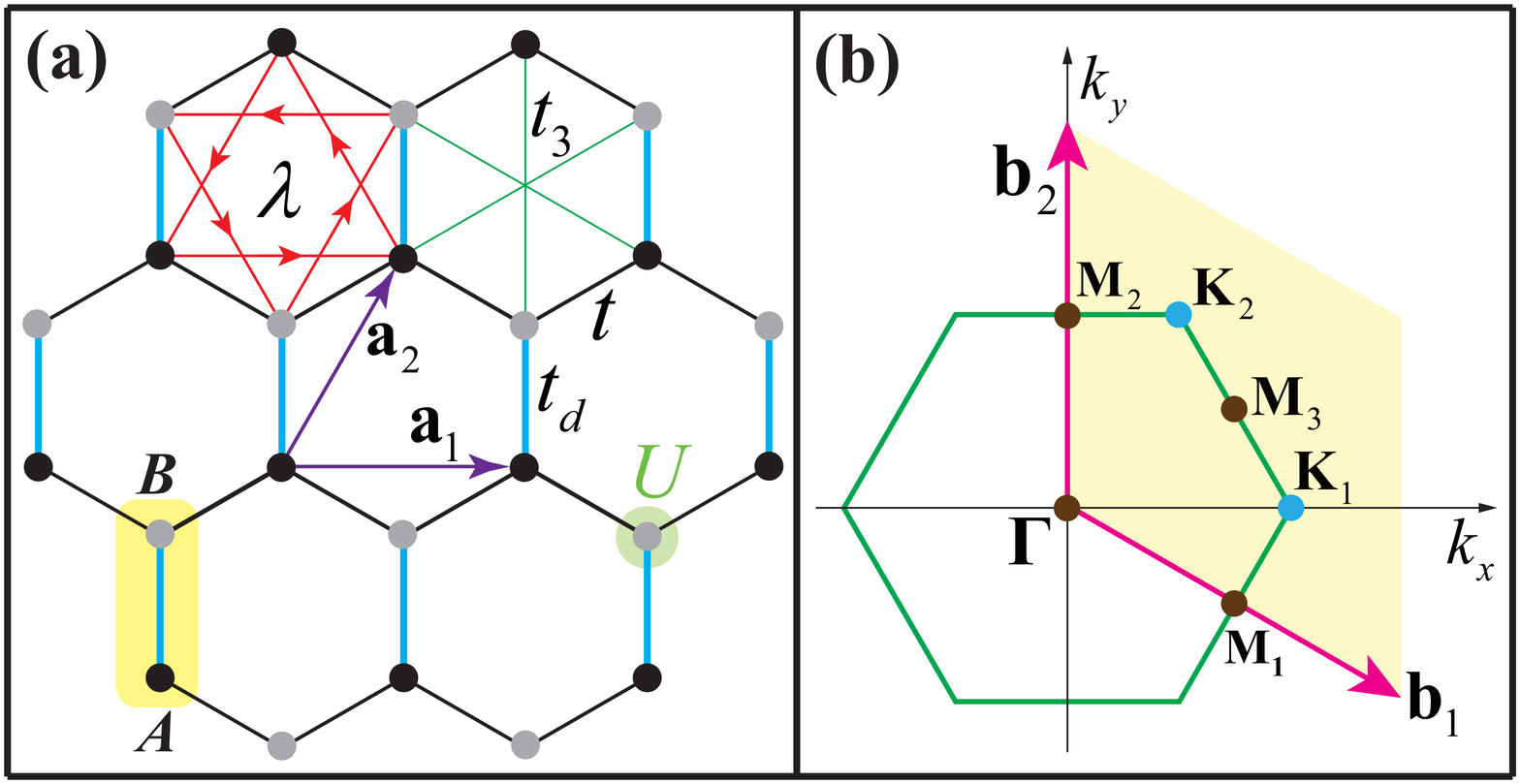}
\caption{\label{fig:KMHLatt}(Color online) (a) Illustration of the GKMH model as in Eq.~(\ref{eq:GKMH}). The unit cell is presented as the yellow shaded rectangle, consisting of A and B sublattices denoted by black and gray dots. The lattice is spanned by primitive vectors $\mathbf{a}_1=(\sqrt{3},0)a$, $\mathbf{a}_2=(1/2,\sqrt{3}/2)a$ with $a$ the lattice constant. The black and blue lines denote nearest-neighbor hopping $t$ and $t_{d}$, while the $\lambda$ term and third-nearest-neighbor hopping $t_{3}$ are represented by red and green lines. The arrows in red lines shows $\nu_{ij}=+1$ for spin-up part. (b) BZ of GKMH model. $\mathbf{K}_1,\mathbf{K}_2$ are Dirac Points, while $\boldsymbol{\kappa}=\mathbf{\Gamma},\mathbf{M}_1,\mathbf{M}_2,\mathbf{M}_3$ are the four TRIM points.}
\end{figure}
For nearest-neighbor (NN) hopping, we have $t_{ij}=t_d$ for NN bonds inside unit cells and $t_{ij}=t$ for the others, as demonstrated in Fig.~\ref{fig:KMHLatt} (a).  The $t_{3}$ term is the third-nearest-neighbor hopping. The fourth term represents spin-orbit coupling ($\lambda$) connecting next-nearest-neighbor sites with a complex (time-reversal symmetric) hopping. The factor $\nu_{ij}=-\nu_{ji}=\pm1$ depends on the orientation of the two nearest-neighbor bonds that the electron moves in going from site $i$ to $j$. The last term describes the on-site Coulomb repulsion. For GKMH model, we set $t$ as energy unit.

\begin{figure}[t]
\centering
\includegraphics[width=\columnwidth]{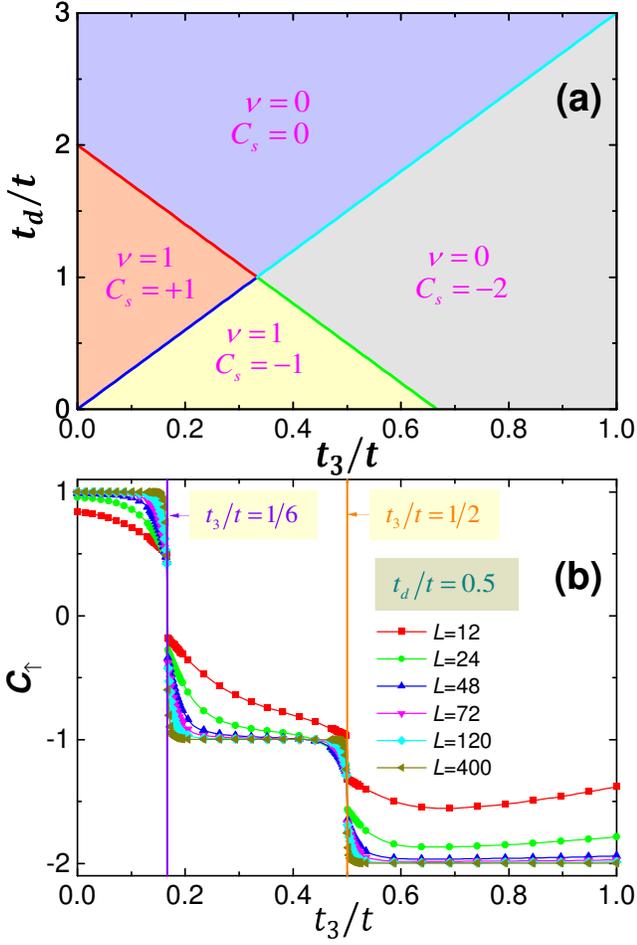}
\caption{\label{fig:GKMHPhase}(Color online) (a) $(t_d/t)-(t_3/t)$ Phase diagram for GKMH model at $U=0$ and arbitrary $\lambda/t>0$. (b) Calculation results of Chern number $C_{\uparrow}$ with $t_d/t=0.5$ and varying $t_3/t$, the linear system sizes are $L=12,24,48,72,120,400$.}
\end{figure}

Due to the $U(1)_{charge}\times U(1)_{spin}\rtimes Z_2^T$ symmetry of the system, GKMH model acquires a $\mathbb{Z}$ classification. Without interaction $(U=0)$, the $(t_d/t)-(t_3/t)$ phase diagram for $\lambda/t>0$, determined from both $Z_2$ invariant $(-1)^\nu$ and Chern Number $C_s$, is shown in Fig.~\ref{fig:GKMHPhase} (a). There are four phases and their phase boundaries in  Fig.~\ref{fig:GKMHPhase} (a) is independent  of the size of $\lambda/t$, as long as $\lambda/t>0$. In Fig.~\ref{fig:GKMHPhase} (b), we show the spin Chern number $C_{\uparrow}$ for fixed $t_d/t=0.5$ and increasing $t_3/t$, calculated from Eq.~(\ref{eq:ChernProject}). Since the system is noninteracting, we have $G_{\sigma}(i\omega=0,\mathbf{k})=-[\mathcal{H}_{\sigma}(\mathbf{k})]^{-1}$ with $\mathcal{H}_{\sigma}(\mathbf{k})$ the $2\times2$ noninteracting Hamiltonian matrix for GKMH model. In Fig.~\ref{fig:GKMHPhase} (b), one can clearly observe that the Chern number $C_s$ converges to its expected quantized value, with increasing lattice size.

With finite interaction strength $U$, we need to measure the single-particle Green's function $G_{\sigma}(\tau,\mathbf{k})$ in QMC simulation and then construct the topological invariants from it. For GKMH model, $G_{\sigma}(\tau,\mathbf{k})$ is a $2\times2$ Hermitian matrix, according to Eq.~(\ref{eq:Relation1}). Due to the presence of spatial inversion symmetry and Eq.~(\ref{eq:Relation2}), we can obtain $[G_{\sigma}(\tau,\mathbf{k})]_{11}=[G_{\sigma}(\tau,-\mathbf{k})]_{22}$ and $[G_{\sigma}(\tau,\mathbf{k})]_{12}=[G_{\sigma}(\tau,-\mathbf{k})]_{21}$. Then at TRIM points $\boldsymbol{\kappa}$, we have $[G_{\sigma}(\tau,\boldsymbol{\kappa})]_{11}=[G_{\sigma}(\tau,\boldsymbol{\kappa})]_{22}$ and $[G_{\sigma}(\tau,\boldsymbol{\kappa})]_{12}=[G_{\sigma}(\tau,\boldsymbol{\kappa})]_{21}$. From Eq.~(\ref{eq:Relation3}), we can get $[G_{\sigma}(\tau,\mathbf{k})]_{pp}=-[G_{\sigma}(-\tau,-\mathbf{k})]_{pp},p=1,2$ and $[G_{\sigma}(\tau,\mathbf{k})]_{12}=[G_{\sigma}(-\tau,-\mathbf{k})]_{21}$. Combining these relations, we obtain the equations applied in numerical calculations as $[G_{\sigma}(-\tau,\mathbf{k})]_{11}=-[G_{\sigma}(\tau,\mathbf{k})]_{22}$, $[G_{\sigma}(-\tau,\mathbf{k})]_{12}=[G_{\sigma}(\tau,\mathbf{k})]_{12}$. At TRIM points, $G_{\sigma}(\tau,\boldsymbol{\kappa})$ is a real symmetric matrix with equal diagonal elements. Based on these considerations, we deduce that there are only two independent matrix elements in $G_{\sigma}(\tau,\mathbf{k})$ as
\begin{eqnarray}
\label{eq:GIw0KMat1}
&&[G_{\sigma}(i\omega=0,\mathbf{k})]_{11} = - [G_{\sigma}(i\omega=0,\mathbf{k})]_{22} \nonumber \\
&&\approx \int_{0}^{+\theta} \Big\{ [G_{\sigma}(\tau,\mathbf{k})]_{11} - [G_{\sigma}(\tau,\mathbf{k})]_{22} \Big\} d\tau,
\end{eqnarray}
and
\begin{eqnarray}
\label{eq:GIw0KMat2}
&&[G_{\sigma}(i\omega=0,\mathbf{k})]_{12} = [G_{\sigma}(i\omega=0,\mathbf{k})]_{21}^\star \nonumber \\
&&\approx 2\int_{0}^{+\theta} [G_{\sigma}(\tau,\mathbf{k})]_{12}d\tau.
\end{eqnarray}
As mentioned above, $\theta$ is the cutoff for $\tau$ in the integral of Eq.~(\ref{eq:Relation2}). Eq.~(\ref{eq:GIw0KMat1}) and Eq.~(\ref{eq:GIw0KMat2}) explicitly show that we only need to measure two elements of $G_{\sigma}(\tau,\mathbf{k})$ matrix with $\tau>0$ at all discrete $\mathbf{k}$ points. At TRIM points, we can obtain the simplified relations as $[G_{\sigma}(i\omega=0,\boldsymbol{\kappa})]_{11}=[G_{\sigma}(i\omega=0,\boldsymbol{\kappa})]_{22}=0$ and $[G_{\sigma}(i\omega=0,\boldsymbol{\kappa})]_{12} = [G_{\sigma}(i\omega=0,\boldsymbol{\kappa})]_{21}$. Namely, at TRIM points we have $G_{\sigma}(i\omega=0,\boldsymbol{\kappa})=Z_{\boldsymbol{\kappa}}\sigma_x$ while $\sigma_x$ is the Pauli Matrix and $Z_{\boldsymbol{\kappa}}$ is some $\boldsymbol{\kappa}$-dependent coefficient~\cite{Lang2013,Hung2013,Hung2014,Meng2014}.

For $Z_2$ invariant, we only deal with $G_{\sigma}(i\omega=0,\boldsymbol{\kappa})$. The matrix representation of inversion symmetry operator $\hat{P}$ for GKMH model is $P=\sigma_x$ for each spin sector. The parity of all unoccupied eigenstates at some TRIM point $\boldsymbol{\kappa}$ is simply $\eta(\boldsymbol{\kappa})=\text{sgn}\{[G_{\sigma}(0,\boldsymbol{\kappa})]_{12}\}$. Then we can get the $Z_2$ invariant simply as $(-1)^\nu=\eta(\boldsymbol{\Gamma})\eta(\mathbf{M}_1)\eta(\mathbf{M}_2)\eta(\mathbf{M}_3)$. As a result, we observe that the $Z_2$ invariant for GKMH model is integer-quantized, free from finite-size effect. On the other hand, to calculate spin Chern number $C_s$, we only need to obtain $[G_{\sigma}(\tau,\mathbf{k})]_{11}$ and $[G_{\sigma}(\tau,\mathbf{k})]_{12}$ in $G_{\sigma}(\tau,\mathbf{k})$ matrix due to its Hermiticity. After that, spin Chern number can be numerically evaluated through applying Eq.~(\ref{eq:ChernProject}).

\begin{figure}[ht!]
\centering
\includegraphics[width=\columnwidth]{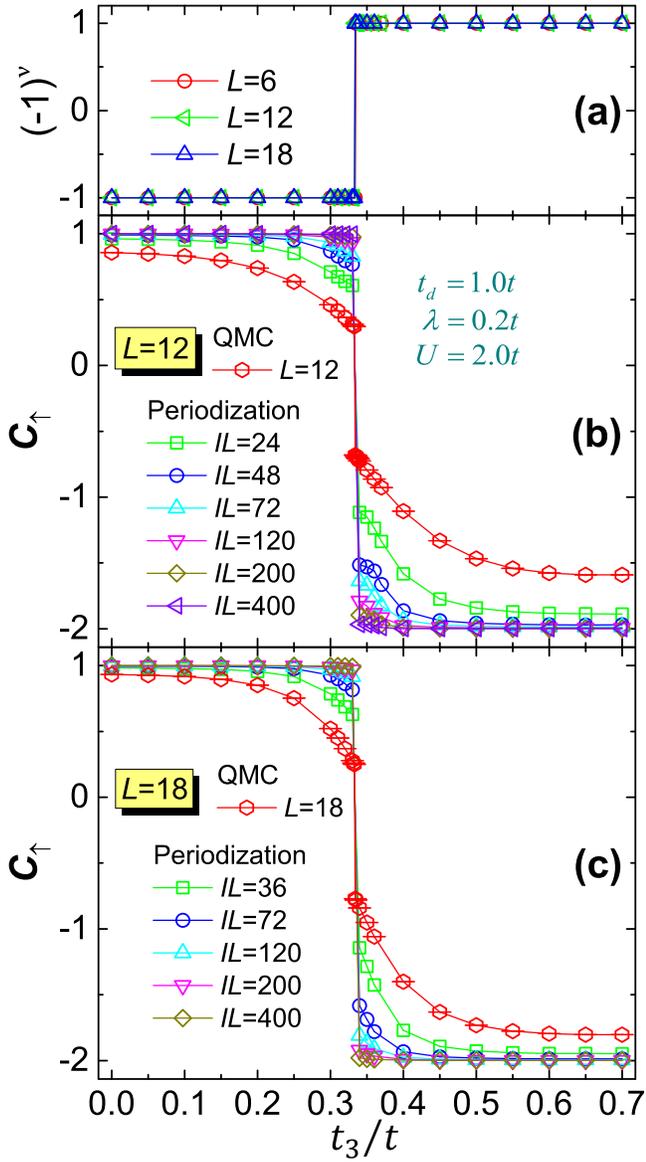}
\caption{\label{fig:Z2ChernT3}(Color online) (a) $Z_2$ invariant $(-1)^\nu$ and (b), (c) Chern number $C_{\uparrow}$ for the $t_3$-driven topological phase transition in GKMH model with $t_d/t=1.0$, $\lambda/t=0.2$ and $U/t=2.0$ from finite-size QMC simulation and the periodization results. (a) $Z_2$ invariant $(-1)^\nu$ is quantized as mentioned in the main text. (b) and (c), the Chern number $C_{\uparrow}$ from finite-size QMC calculation indicated by the red open hexagon with error bar acquires a drop with finite value at the transition point, which can be taken as signature of topological phase transition,  but the $C_{\uparrow}$ itself is not quantized before or after the phase transition, due to the finite-size effect. After the periodization with the QMC data in $L=12$ (b) and $L=18$ (c) systems, the $C_{\uparrow}$ converge to the ideal quantized integers, where $IL$ stands for interpolation lattice size used in the {\it periodization} process.}
\end{figure}

To apply our numerical calculation scheme for topological invariants, we choose two paths in the phase diagram of Fig.~\ref{fig:GKMHPhase} (a). First, starting from $t_d/t=1.0$, $\lambda/t=0.2$ and $U/t=2.0$, we calculate the topological invariants to  monitor the $t_3$-driven topological phase transition. Second, we choose $t_3/t=0$, $\lambda/t=0.2$, $U/t=2.0$ and calculate the topological invariants to monitor the $t_d$-driven topological phase transition. The interaction is chosen to be at a small value $U/t=2.0$ to avoid the appearance of antiferromagnetic state~\cite{Meng2014,HQWu2015}. Thus, there is no spontaneous symmetry breaking across these two topological phase transitions driven by hopping parameters in the interacting GKMH model.


For the $t_3$-driven topological phase transition, the results of both $Z_2$ invariant and Chern number $C_{\uparrow}$ (equal to spin Chern number $C_s$) from QMC simulations with finite size $L=6,12,18$ are shown in Fig.~\ref{fig:Z2ChernT3} (a), (b) and (c), at $t_{d}/t=1.0, \lambda/t=0.2$ and $U/t=2.0$. At $U=0$, the GKMH model experiences a topological phase transition at $t_3/t=1/3$ from $C_s=+1$ to $C_s=-2$ as indicated in Fig.~\ref{fig:GKMHPhase} (a). At $U/t=2.0$, the quantized $Z_2$ invariant in Fig.~\ref{fig:Z2ChernT3} (a) demonstrates that the topological phase transition point changes little. Detailed calculations of parities at all TRIM points show that the parities at $\mathbf{M}_1,\mathbf{M}_2,\mathbf{M}_3$ change across the phase transition, which is related to the fact that the single-particle gap closes at $\mathbf{M}_1,\mathbf{M}_2,\mathbf{M}_3$ points~\cite{Meng2014}. As for the Chern number from finite-size QMC calculation denoted by by the red open hexagon symbols with error bar in Fig.~\ref{fig:Z2ChernT3} (b) and (c), we can observe drop with finite values across the phase transition, the position coincide with that in $Z_2$ invariant. Combining these results, the phase transition point is $t_3/t\approx 0.334\thicksim~0.335$ for $L=6$ system and $t_3/t\approx 0.333\thicksim0.334$ for $L=12,18$ system.

The problem about the results of Chern number in Fig.~\ref{fig:Z2ChernT3} (b) and (c) is that they are not integer-quantized, though the convergence with increasing $L$ can be observed. 
To solve this problem, the periodization method described in Sec.~\ref{sec:Periodization} is applied based on the results of Chern number from QMC simulations of $L=12,18$. As shown in Fig.~\ref{fig:Z2ChernT3} (b) and (c), one indeed observes the gradual convergence of Chern number $C_{\uparrow}$ when the interpolation lattice size $IL$ increases. In fact, for $IL=120$, the Chern number $C_{\uparrow}$ is almost ideally quantized, which demonstrates that the periodization method works very well.

\begin{figure}[ht!]
\centering
\includegraphics[width=\columnwidth]{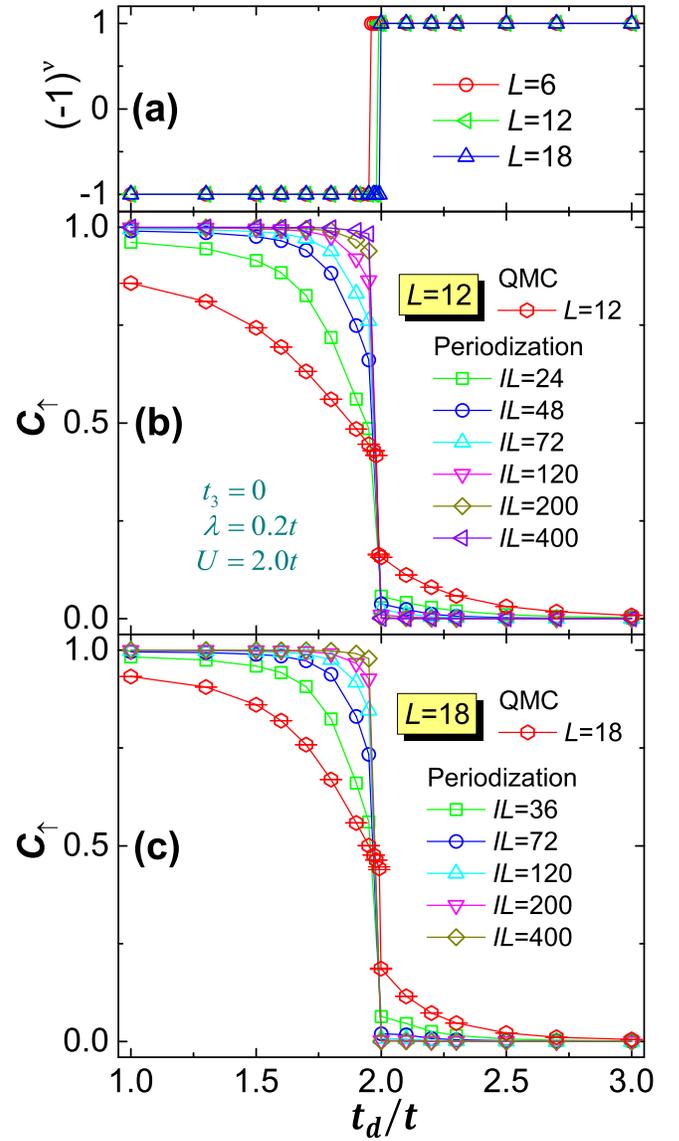}
\caption{\label{fig:Z2ChernTd}(Color online) (a) $Z_2$ invariant $(-1)^\nu$ and (b), (c) Chern number $C_{\uparrow}$ for the $t_d$-driven topological phase transition in GKMH model with $t_3/t=0$, $\lambda/t=0.2$ and $U/t=2.0$ from finite-size QMC simulation and periodization results. (a) $Z_2$ invariant is integer-quantized and there is a gradual shifting of the transition point from $t_d/t\approx1.955$ in $L=6$ system to $t_d/t\approx1.995$ in $L=18$ system. (b) and (c), the finite drop in Chern number from finite-size QMC calculation with $L=12$ and $L=18$ (denoted by red open hexagon with error bar) can be observed clearly.  The periodization results of Chern number $C_{\uparrow}$, with the QMC data in $L=12$ (b) and $L=18$ (c) systems, are ideally quantized.}
\end{figure}


For the $t_d$-driven topological phase transition, the results of both $Z_2$ invariant and Chern number $C_{\uparrow}$ (equal to spin Chern number $C_s$) from finite-size QMC simulations are shown in Fig.~\ref{fig:Z2ChernTd}, at $t_3/t=0$, $\lambda/t=0.2$ and $U/t=2.0$. The noninteracting GKMH model obtains a topological phase transition at $t_d/t=2.0$ for $t_3/t=0$, as long as $\lambda/t>0$. From the results in Fig.~\ref{fig:Z2ChernTd}, one sees the weak interaction $U/t=2.0$ only give a small shift of the topological phase transition point. Across this $t_d$-driven phase transition, both the parity change and single-particle gap close happen only at the $\mathbf{M}_2$ point, due to the anisotropy~\cite{Hung2013,Hung2014,Meng2014} introduced by $t_d$. From the integer-quantized $Z_2$ invariant, the phase transition point only possesses a small shift in $t_d/t$, from $L=6$ system to $L=18$ system, in Fig.~\ref{fig:Z2ChernTd} (a). By means of the {\it periodization} process, with the QMC simulation results of $L=12$ and $18$, we obtain the perfect, quantized Chern number $C_{\uparrow}$ results in Fig.~\ref{fig:Z2ChernTd} (b) and (c).


\subsection{Cluster Kane-Mele-Hubbard model}
\label{sec:ClusterKMH}
\begin{figure}[t]
\centering
\includegraphics[width=0.9\columnwidth]{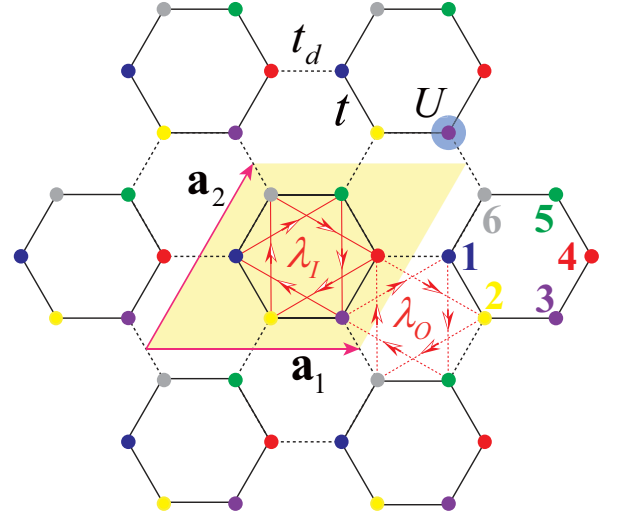}
\caption{\label{fig:ClusterKMH}(Color online) Illustration of the CKMH model in Eq.~(\ref{eq:CKMH}). The yellow shaded region shows the six-site unit cell with primitive lattice vectors $\mathbf{a}_1=(\sqrt{3},0)a$, $\mathbf{a}_2=(1/2,\sqrt{3}/2)a$ with the nearest-neighbor bond length $a/\sqrt{3}$. The six sublattices $1,2,3,4,5,6$ are shown in different color. The black solid and black dotted lines indicate the nearest-neighbor hopping term inside ($t$) and between ($t_{d}$) unit cells. The red solid and red dotted lines represent SOC terms inside ($\lambda_{I}$) and between ($\lambda_{O}$) unit cells. The sign choice for SOC hopping is the same as that in Fig.~\ref{fig:KMHLatt} (a). The on-site Coulomb repulsion $U$ is shown by the blue shaded circle.}
\end{figure}

The cluster Kane-Mele-Hubbard model~\cite{Grandi2015a,Wu2012} (CKMH) has six honeycomb lattice sites as one unit cell, the Hamiltonian is given as follows
\begin{eqnarray}
\hat{H} =&& -\sum_{\langle ij\rangle\sigma}t_{ij}(c^{\dagger}_{i\sigma}c_{j\sigma} + c^{\dagger}_{j\sigma}c_{i\sigma})  \nonumber\\
&& +i\lambda_{I}\sum_{\langle\!\langle ij \rangle\!\rangle \alpha\beta}v_{ij}(c^{\dagger}_{i\alpha}\sigma^{z}_{\alpha\beta}c_{j\beta}
-c^{\dagger}_{j\beta}\sigma^{z}_{\beta\alpha}c_{i\alpha}) \nonumber\\
&& +i\lambda_{O}\sum_{\langle\!\langle ij \rangle\!\rangle \alpha\beta}v_{ij}(c^{\dagger}_{i\alpha}\sigma^{z}_{\alpha\beta}c_{j\beta}
-c^{\dagger}_{j\beta}\sigma^{z}_{\beta\alpha}c_{i\alpha}) \nonumber\\
&& +\frac{U}{2}\sum_{i}(n_{i\uparrow}+n_{i\downarrow}-1)^2\;.
\label{eq:CKMH}
\end{eqnarray}
For nearest-neighbor (NN) hopping, we have $t_{ij}=t$ for NN bonds inside unit cells and $t_{ij}=t_d$ for those connecting the six-site unit cells, as demonstrated in Fig.~\ref{fig:ClusterKMH} (a). The amplitudes for SOC term inside a unit cell and between different unit cells are $\lambda_I$ and $\lambda_O$, respectively. $U$ is the on-site Coulomb repulsion. Similar to the GKMH model, the $U(1)_{change}\times U(1)_{spin}\rtimes Z_2^T$ symmetry is also preserved in CKMH model, which results in $\mathbb{Z}$ classification. Besides, both spatial inversion symmetry and particle-hole symmetry are also present in CKMH model. Notice that the CKMH model has 6-site unit cell, in this section, the linear system size $L$ in finite size QMC simulation actually corresponds to that of the a 6-site unit cell, i.e., the total lattice sites are $6\times L\times L$. Such large unit cell greatly increases the QMC simulation efforts of CKMH model comparing to that of the GKMH model in previous session, where the total lattice site is only $2\times L\times L$ (the computation efforts of QMC scale to the third power of the total lattice sites).

\begin{figure}[t]
\centering
\includegraphics[width=\columnwidth]{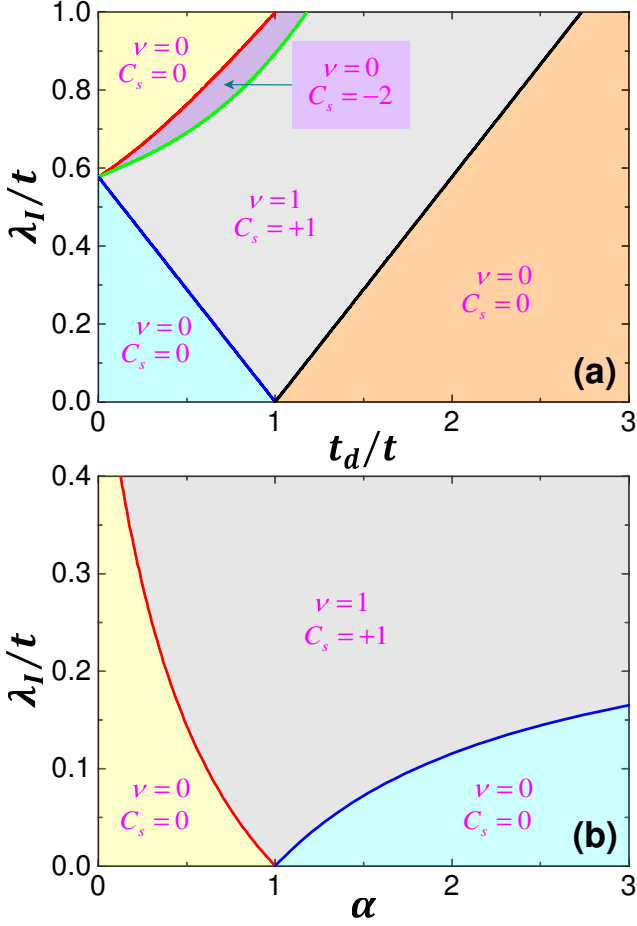}
\caption{\label{fig:ClusterKMHPhase}(Color online) Phase diagrams for noninteracting CKMH model under two different parameter sets. (a) $(t_d/t)-(\lambda_I/t)$ phase diagram with $\lambda_O=0$; (b) $\alpha-(\lambda_I/t)$ phase diagram with $\alpha=t_d/t=\lambda_O/\lambda_I$.}
\end{figure}

Since there are three independent parameters $t_d,\lambda_I,\lambda_O$, even the phase diagram for noninteracting CKMH model is already interesting. To simplify the presentation, we only demonstrate results on two special cases. First, we set $\lambda_O=0$, the $(t_d/t)-(\lambda_I/t)$ phase diagram of this case is shown in Fig.~\ref{fig:ClusterKMHPhase} (a). We can observe that the QSHI phases exist in the middle region of the $(t_d/t)-(\lambda_I/t)$ phase diagram, with different spin Chern numbers $C_s=+1$ and $C_s=-2$, as a function of $\lambda_{I}/t$. The presence of the $C_s=-2$ phase is unexpected and very interesting, since the Chern number is changed by 3 when going from $C_{s}=1$ to $C_{s}=-2$, further increase $\lambda_{I}/t$, QSHI phase is destroyed. Second, we keep all three $t_d,\lambda_I,\lambda_O$ parameters finite and introduce a ratio of hopping $\alpha=t_d/t=\lambda_O/\lambda_I$. The $\alpha-(\lambda_I/t)$ phase diagram at $U=0$ is presented in Fig.~\ref{fig:ClusterKMHPhase} (b). In the plotted region of $\alpha$ and $t_d/t$, three phases with one non-trivial in $C_s$ are found.

With interaction, we apply the {\it periodization} process in Sec.~\ref{sec:Periodization} to calculate the topological invariants. Before presenting the data, let's discuss the structure of $G_{\sigma}(\tau,\mathbf{k})$ and $G_{\sigma}(i\omega=0,\mathbf{k})$ for CKMH model as it is quite complicated. First, both $G_{\sigma}(\tau,\mathbf{k})$ and $G_{\sigma}(i\omega=0,\mathbf{k})$ are $6\times6$ Hermitian matrices for CKMH model. Second, using spatial inversion and particle-hole symmetries, we can obtain useful relations among the matrix elements of $G_{\sigma}(\tau,\mathbf{k})$. Combining these two symmetry properties, $G_{\sigma}(\tau,\mathbf{k})$ and $G_{\sigma}(-\tau,\mathbf{k})$ are explicitly related and we only need to calculate the $G_{\sigma}(\tau,\mathbf{k})$ data with $\tau>0$. The detailed analysis is presented in Appendix~\ref{sec:appendix_d}. The $G_{\sigma}(i\omega=0,\mathbf{k})$ matrix for the CKMH model only has 12 independent matrix elements and can be expressed as
\begin{eqnarray}
\label{eq:J22Equation}
G_{\sigma}(0,\mathbf{k})=\left(                 
  \begin{array}{cccccc}   
    A_1         &   A_4           &   A_5            &   A_6   &   A_7           &   A_8            \\
    A_4^\star   &   A_2           &   A_9            &  -A_7   &   A_{10}        &   A_{11}         \\
    A_5^\star   &   A_9^\star     &   A_3            &   A_8   &  -A_{11}        &   A_{12}         \\
    A_6^\star   &  -A_7^\star     &   A_8^\star      &  -A_1   &   A_4^{\star}   &  -A_5^{\star}    \\
    A_7^\star   &   A_{10}^\star  &  -A_{11}^\star   &   A_4   &  -A_2           &   A_9^{\star}    \\
    A_8^\star   &   A_{11}^\star  &   A_{12}^\star   &  -A_5   &   A_9           &  -A_3
  \end{array}
    \right) , \hspace{0.4cm}
\end{eqnarray}
where $A_1,A_2,A_3$ are real numbers and $A_i,i=4,\cdots,12$ are complex numbers. At TRIM points, Eq.~(\ref{eq:J22Equation}) can be further simplified to only 9 independent matrix elements and $G_{\sigma}(i\omega=0,\boldsymbol{\kappa})$ matrix obtains the following matrix structure as
\begin{eqnarray}
\label{eq:J42Equation}
&&G_{\sigma}(0,\boldsymbol{\kappa})= \nonumber \\
&& \left(                 
  \begin{array}{cccccc}   
    0     &   B_4     &   iB_5     &   B_6   &   iB_7          &   B_8            \\
    B_4   &   0       &   B_9      &  -iB_7  &   B_{10}        &   iB_{11}         \\
   -iB_5  &   B_9     &   0        &   B_8   &  -iB_{11}       &   B_{12}         \\
    B_6   &  iB_7     &   B_8      &  0      &   B_4           &  iB_5    \\
   -iB_7  &   B_{10}  &  iB_{11}   &   B_4   &    0            &   B_9    \\
    B_8   & -iB_{11}  &   B_{12}   & -iB_5   &   B_9           &  0
  \end{array}
    \right). \hspace{0.5cm}
\end{eqnarray}
In Eq.~(\ref{eq:J42Equation}), $B_i,i=4,\cdots,12$ are real numbers. The explicit expressions for both $A_i$ and $B_i$ are given in Appendix~\ref{sec:appendix_d}.

For numerical evaluation of $Z_2$ invariant in CKMH model, we adopt the $G_{\sigma}(i\omega=0,\boldsymbol{\kappa})$ matrix in Eq.~(\ref{eq:J42Equation}) for TRIM points. The matrix representation of spatial inversion symmetry operator for each spin sector is
\begin{eqnarray}
\label{eq:J43Equation}
P =\left(                 
  \begin{array}{cccccc}   
    0     &   0     &   0     &   1  &   0          &   0     \\
    0     &   0     &   0     &   0  &   1          &   0     \\
    0     &   0     &   0     &   0  &   0          &   1     \\
    1     &   0     &   0     &   0  &   0          &   0     \\
    0     &   1     &   0     &   0  &   0          &   0     \\
    0     &   0     &   1     &   0  &   0          &   0
  \end{array}
    \right).
\end{eqnarray}
To calculate the $Z_2$ invariant, we only need to diagonalize the $G_{\sigma}(i\omega=0,\boldsymbol{\kappa})$ matrix with $\boldsymbol{\kappa}=\boldsymbol{\Gamma},\mathbf{M}_1,\mathbf{M}_2,\mathbf{M}_3$, and then calculate the parities at these TRIM points by Eq.~(\ref{eq:Z2Int}). To calculate spin Chern number, we only need to adopt the matrix structure of $G_{\sigma}(i\omega=0,\mathbf{k})$ in Eq.~(\ref{eq:J22Equation}) and apply Eq.~(\ref{eq:ChernProject}), first with the finite-size QMC data and then with {\it periodization} process.

\begin{figure}[ht!]
\centering
\includegraphics[width=\columnwidth]{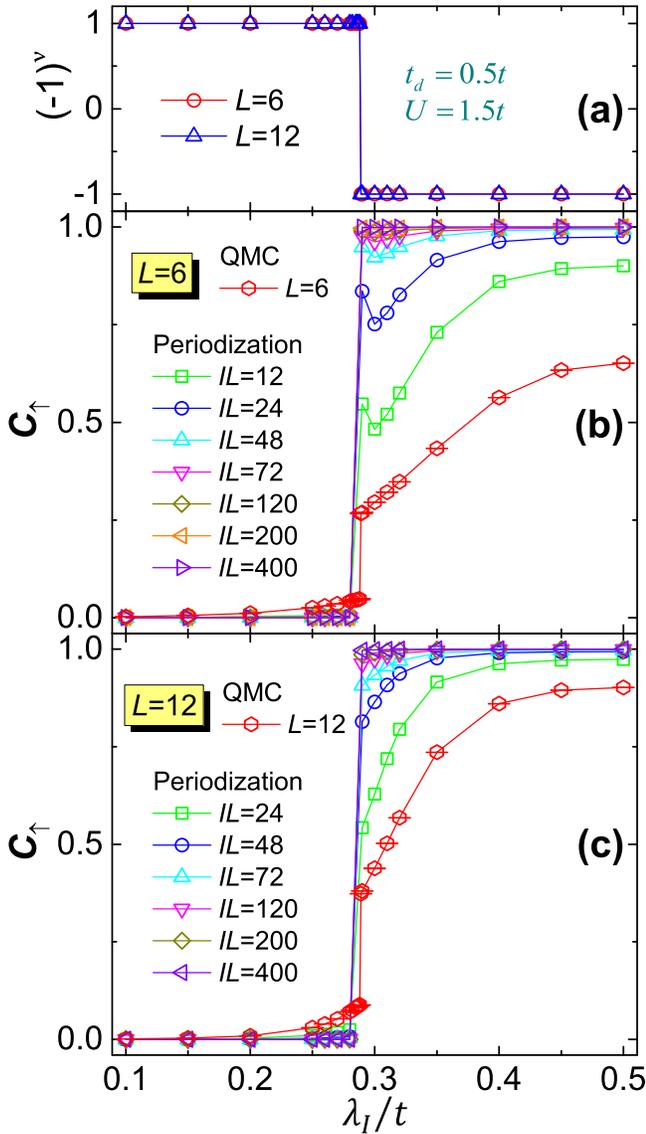}
\caption{\label{fig:CKMH1_Z2Chern}(Color online) (a) Z2 invariant $(-1)^\nu$ and (b), (c) Chern number $C_{\uparrow}$ for the $\lambda_I$-driven topological phase transition with $\lambda_{O}=0$, $t_d/t=0.5$ and $U/t=1.5$ from finite-size calculations by QMC simulation (denoted by red open hexagon with error bar) and periodization. The drop of integer-valued $Z_2$ invariant and the jump of Chern Number $C_{\uparrow}$ can be taken as signature of topological phase transition. In (b) and (c), with large interpolation lattice size $IL$ from finite QMC data with $L=6$ and $12$, the Chern number $C_{\uparrow}$ reaches its quantized value.}
\end{figure}

We concentrate on two independent paths in the parameter space of CKMH model. First, we choose $\lambda_{O}=0$, $t_d/t=0.5$ and $U/t=1.5$ and study the $\lambda_I$-driven topological phase transition in the interacting CKMH model. For $U=0$, the transition point for this $\lambda_I$-driven topological phase transition from $C_s=0$ to $C_s=+1$ is at $\lambda_I/t\approx0.289$. Second, we set $\alpha=1.8,U/t=2.0$ in the CKMH model and study the $\lambda_I$-driven topological phase transition, similarly, there is also a $\lambda_I$-driven topological phase transition from $C_s=0$ to $C_s=+1$ at $\lambda_I/t\approx0.1$ for $U=0$. In the following, we calculate $Z_2$ invariant $(-1)^\nu$ and spin Chern number $C_s$ for $\lambda_I/t$ parameter across these two phase transitions, to demonstrate that our {\it periodization} process works for CKMH model as well.


Fig.~\ref{fig:CKMH1_Z2Chern} shows the $Z_2$ invariant $(-1)^\nu$ and Chern number $C_{\uparrow}$ (equal to spin Chern number $C_s$), calculated from finite-size QMC simulations of $L=6,12$ systems (both denoted by red open hexagon with error bar) for the $\lambda_{O}=0$, $t_d/t=0.5$ and $U/t=1.5$ case. One can observe that the $Z_2$ invariant is exactly integer-quantized to $\pm1$ (Fig.~\ref{fig:CKMH1_Z2Chern} (a)). Such integer quantization in finite-size system is due to the spatial inversion and particle-hole symmetries of the CKMH model. These two symmetries result in the special matrix structure of $G_{\sigma}(i\omega=0,\boldsymbol{\kappa})$ in Eq.~(\ref{eq:J42Equation}), which is sufficient to guarantee that the parities at TRIM points are exactly $\pm1$. As shown in Fig.~\ref{fig:CKMH1_Z2Chern} (a), (b) and (c), for both $L=6$ and $L=12$, the $\lambda_I$-driven topological phase transition point is at $\lambda_I/t=0.288\thicksim0.289$ from the drop in $Z_2$ invariant $(-1)^\nu$ and the jump in Chern number $C_{\uparrow}$. Across the topological phase transition, both single-particle gap close and the parity change all happen at $\boldsymbol{\Gamma}$ point. As for the Chern number $C_{\uparrow}$ from finite size $L=6$ and $L=12$ QMC calculation (Fig.~\ref{fig:CKMH1_Z2Chern} (b) and (c)), they are still far from the ideal quantized result due to the finite-size effect, although the trend of convergence with increasing system size is present.

\begin{figure}[ht!]
\centering
\includegraphics[width=\columnwidth]{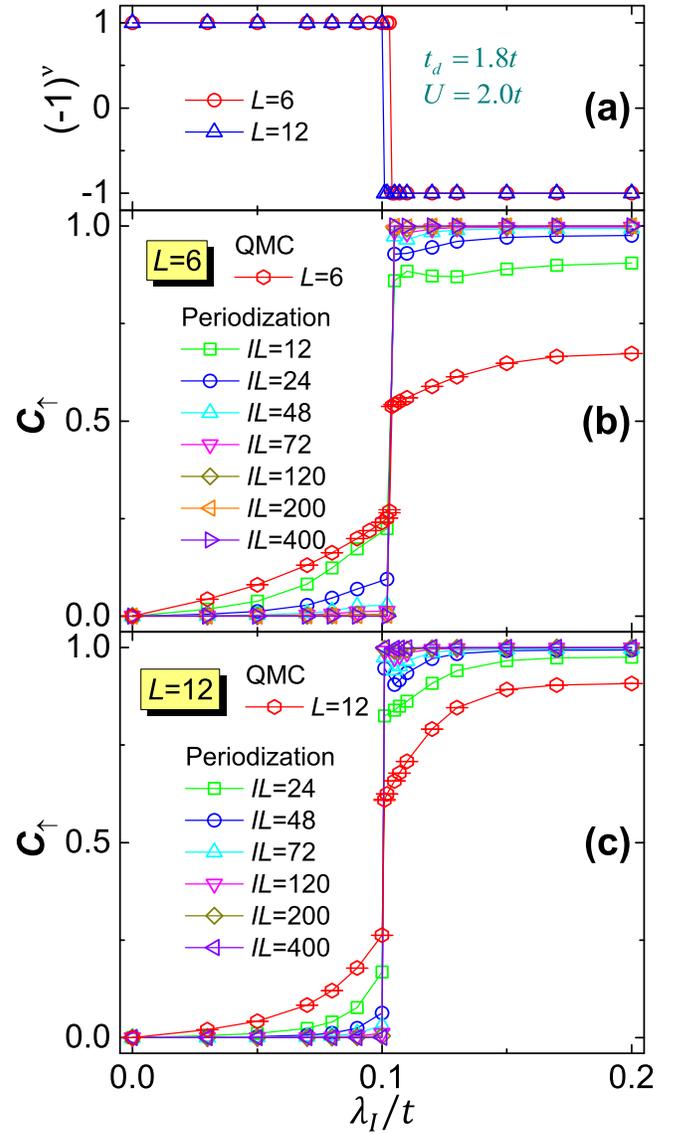}
\caption{\label{fig:CKMH2_Z2Chern}(Color online) (a) Z2 invariant $(-1)^\nu$ and (b), (C) Chern number $C_{\uparrow}$ for the $\lambda_I$-driven topological phase transition with $\alpha=1.8$ and $U/t=2.0$ from finite-size calculations by QMC (denoted by red open hexagon with error bar) and periodization. The drop of integer-valued $Z_2$ invariant and the jump of Chern number $C_{\uparrow}$ can be taken as signature of topological phase transition. In (b) and (c), with large $IL$ in periodization, the Chern number $C_{\uparrow}$ reaches its quantized value.}
\end{figure}

The calculation results of Chern number $C_{\uparrow}$ by the periodization method for the $\lambda_I$-driven topological phase transition is presented in Fig.~\ref{fig:CKMH1_Z2Chern} (b) and (c), from the QMC data in $L=6$ and $12$ systems. We can observe that the periodization method can give good results of quantized integer for $C_{\uparrow}$. The nonmonotonic behavior in $IL=12,24,48$ close to the transition in Fig.~\ref{fig:CKMH1_Z2Chern} (b) is due to the inappropriate $\tau$ cutoff in calculating Eq.~(\ref{eq:NewFourier3}), as mentioned in Sec.~\ref{sec:Numerical}. This behavior is absent for large enough $IL$s. Through this periodization method, a sharp topological phase transition from $C_s=0$ to $C_s=+1$ can be clearly seen.


Fig.~\ref{fig:CKMH2_Z2Chern} shows the results of $Z_2$ invariant $(-1)^\nu$ (Fig.~\ref{fig:CKMH2_Z2Chern} (a)) and Chern number $C_{\uparrow}$ (Fig.~\ref{fig:CKMH2_Z2Chern} (b), (c)) for the case of $\lambda_I$-driven topological phase transition with $\alpha=1.8$, $U/t=2$, from finite-size QMC simulations of $L=6$, $12$ systems and periodization. The sharp drop of integer-valued $Z_2$ invariant $(-1)^\nu$ defines the topological phase transition, at $\lambda_I/t=0.103\thicksim0.104$ for $L=6$ system and $\lambda_I/t=0.100\thicksim0.101$ for $L=12$ system. The positions for the finite value jump of Chern number $C_{\uparrow}$ in $L=6$ and $L=12$ systems are consistent with those of $Z_2$ invariant $(-1)^\nu$. Across the topological phase transition, both the single-particle gap close and the parity change happen at $\boldsymbol{\Gamma}$ point. Still, $C_{\uparrow}$ from QMC are not quantized integer value according to the results in Fig.~\ref{fig:CKMH2_Z2Chern} (b) and (c).

The calculation results of Chern number $C_{\uparrow}$ after the periodization for the $\lambda_I$-driven topological phase transition is presented in Fig.~\ref{fig:CKMH2_Z2Chern} (b) and (c), from the QMC data in $L=6$ and $12$ systems. With large enough $IL$, the integer-valued Chern number $C_{\uparrow}$ can be obtained. Again, very close to the transition point, the nonmonotonic behavior of $C_{\uparrow}$ from periodization method appears both in Fig.~\ref{fig:CKMH2_Z2Chern} (b) and (c), when $IL$ is not large, but it disappears after we increase $IL$.


\section{Conclusion}
\label{sec:Conclusion}

To conclude, we provide a toolkit to calculate the topological invariants constructed from zero-frequency single-particle Green's function for interacting TIs. All the important numerical details are carefully documented, hence serves the purpose of demystifying the numerical evaluation of $Z_2$ invariant and spin Chern number for interacting TIs. Most importantly, we introduce a {\it periodization} process to eliminate the finite-size effect on spin Chern number and obtain quantized topological invariants from finite-size QMC simulations, which renders the topological phases well-defined.

To demonstrate the power of our calculation scheme, especially the {\it periodization} process, both the topological phases and topological phase transitions in two interacting TI models, namely, the Generalized Kane-Mele-Hubbard model and the cluster Kane-Mele-Hubbard model, are identified by numerical evaluation of topological invariants. The results show that the numerical scheme works well in capturing the topological phases and their phase transitions driven by one-body model parameter. The $Z_2$ invariant are already integer-quantized by applying the symmetry properties of the studied system during data process. With {\it periodization} process, the integer-quantized spin Chern number is also achieved with QMC simulations in systems with very small size. Through these calculations of topological invariants, we can also determine the topological phase transition points accurately, at least more accurate than those from the gap extrapolations.

The present work demonstrates that the numerical evaluation scheme, especially the {\it periodization} process, of topological invariants for interacting TIs works well, in distinguishing topologically phases and identifying their phase transitions driven by the one-body model parameters. In paper (II) of this series, we shall apply the numerical evaluation scheme developed here to wider classes of models of interacting TIs, in which certain limitation of constructing topological invariants from single-particle Green's function is manifested in a very interesting manner, calling for more versatile technique to diagnose the interaction-driven topological phase transitions in interacting TIs.

\begin{acknowledgments}
We thank Ning-Hua Tong, Lei Wang, Yi-Zhuang You, Cenke Xu, Liang Fu and Xi Dai for inspiring discussions, and in particular we acknowledge Zhong Wang for helpful comments on the Green's function formalism and the manuscript. The numerical calculations were carried out at the Physical Laboratory of High Performance Computing in Renmin University of China as well as the National Supercomputer Center in TianJin on the platform Tianhe-1A. YYH, HQW and ZYL acknowledge support from National Natural Science Foundation of China (Grant Nos. 11474356 and 11190024) and National Program for Basic Research of MOST of China (Grant No. 2011CBA00112). ZYM is supported by the National Natural Science Foundation of China (Grant Nos. 11421092 and 11574359) and the National Thousand-Young-Talents Program of China, and acknowledges the hospitality of the KITP at the University of California, Santa Barbara, where part of this work is completed.
\end{acknowledgments}

\newpage
\onecolumngrid

\appendix

\section{Detailed implementation of Eq.~(\ref{eq:ChernProject}) for honeycomb lattice}
\label{sec:appendix_a}
In this appendix, we present the numerical implementation of Eq.~(\ref{eq:ChernProject}) on honeycomb lattice, provided that we have already obtained the zero-frequency single-particle Green's function data $G_{\sigma}(i\omega=0,\mathbf{k})$. The reason for having this discussion is that the primitive vectors for honeycomb lattice in real $(\mathbf{a}_{1}, \mathbf{a}_{2})$ and momentum space $(\mathbf{b}_{1},\mathbf{b}_{2})$ (see Fig.~\ref{fig:KMHLatt} (a), (b)) are not along the $(x, y)$ or $(k_{x}, k_{y})$ direction of the Cartesian coordinate. Hence, both the integral over honeycomb lattice BZ and the derivatives over $(k_x, k_y)$ in Eq.~(\ref{eq:ChernProject}) can be Jacobian transformed from those on the honeycomb lattice BZ to that on a square lattice as
\begin{eqnarray}
\label{eq:JacobiTrans}
\left\{\begin{array}{ll}
    q_u=k_x  \\
    q_v=\frac{k_x+\sqrt{3}k_y}{2}
\end{array}
\right. \hspace*{0.8cm}
\left\{\begin{array}{ll}
    0\le q_u \le \frac{2\pi}{\sqrt{3}a}  \\
    0\le q_v \le \frac{2\pi}{\sqrt{3}a}
\end{array}
\right. \hspace*{0.7cm}  \Longrightarrow \hspace{0.7cm}
k_x = q_u \hspace{0.7cm} k_y = \frac{-q_u+2q_v}{\sqrt{3}}.
\end{eqnarray}
The transformed BZ is indeed a square one, and we can rewrite Eq.~(\ref{eq:ChernProject}) by substituting Eq.~(\ref{eq:JacobiTrans}) as
\begin{eqnarray}
\label{eq:SquareProject}
\mathcal{C}&=&\frac{1}{2\pi i}\int_0^{2\pi/\sqrt{3}a}dq_u\int_0^{2\pi/\sqrt{3}a}dq_v\cdot \text{ Tr}\Big\{Q(q_u,q_v)\Big[\partial_{q_u}Q(q_u,q_v)\partial_{q_v}Q(q_u,q_v)-\partial_{q_v}Q(q_u,q_v)\partial_{q_u}Q(q_u,q_v)\Big]\Big\},
\end{eqnarray}
where $Q(q_u,q_v)=P(q_u,(-q_u+2q_v)/\sqrt{3})$ and $P(k_x,k_y)$ is the projection matrix defined in Eq.~(\ref{eq:ProjectOperator}). Comparing Eq.~(\ref{eq:ChernProject}) and Eq.~(\ref{eq:SquareProject}), one observes the Jacobian transformation does not change the form of the formula.


QMC simulates finite size system, so the integration and derivation in Eq.~(\ref{eq:SquareProject}) are discretized, we set $Q_{i,j}=Q(q_{u,i},q_{v,j})$ and $(q_{u,i},q_{v,j})=(2i\pi/\sqrt{3}L_1a, 2j\pi/\sqrt{3}L_2a)$ with $i\in[0,L_1],j\in[0,L_2]$, 
then we have the expressions for finite difference as
\begin{eqnarray}
\label{eq:SquareFiniteDiff}
\partial_{q_u}Q(q_u,q_v)|_{\mathbf{q}=(q_{u,i},q_{v,j})} = \frac{Q_{i+1,j}-Q_{i-1,j}}{2\delta_{q_u}} \hspace{1.0cm}
\partial_{q_v}Q(q_u,q_v)|_{\mathbf{q}=(q_{u,i},q_{v,j})} = \frac{Q_{i,j+1}-Q_{i,j+1}}{2\delta_{q_v}},
\end{eqnarray}
where $\delta_{q_u}=2\pi/\sqrt{3}L_1a,\delta_{q_v}=2\pi/\sqrt{3}L_2a$. 
Due to the periodic boundary condition, we have $Q_{0,j}=Q_{L_1,j}, Q_{L_1+1,j}=Q_{1,j}$ and $Q_{i,0}=Q_{i,L_2}, Q_{i,L_2+1}=Q_{i,1}$. Based on Eq.~(\ref{eq:SquareFiniteDiff}), we arrive at the expression for the integrand in Eq.~(\ref{eq:SquareProject}) as
\begin{eqnarray}
\label{eq:CSImple1}
\text{Tr}\Big\{&&Q(q_u,q_v)\Big(\partial_{q_u}Q(q_u,q_v)\partial_{q_v}Q(q_u,q_v)-\partial_{q_v}Q(q_u,q_v)\partial_{q_u}Q(q_u,q_v)\Big)\Big\} \\ \nonumber
&&=\frac{1}{4\delta_{q_u}\delta_{q_v}}\text{Tr}\Big\{ Q_{i,j}\Big( [Q_{i+1,j},Q_{i,j+1}]+[Q_{i,j+1},Q_{i-1,j}]+[Q_{i-1,j},Q_{i,j-1}]+[Q_{i,j-1},Q_{i+1,j}] \Big) \Big\}.
\end{eqnarray}
Simultaneously, the integral over the square BZ in Eq.~(\ref{eq:SquareProject}) is changed into summation over discrete wavevector points as
\begin{equation}
\label{eq:CSImple2}
\int_{\mathbf{k}\in BZ} f(\vec{k})d^2 \mathbf{k}=\frac{\Omega_{BZ}}{L_{1}L_{2}}\sum_{\mathbf{k}\in BZ}f(\mathbf{k}),
\end{equation}
where $L_1 L_2$ is number of unit cells for the finite size system, and $\Omega_{BZ}=4\pi^2/3a$ is the volume of BZ. 
For a finite-size system, the result of the summation will deviate from the expected quantized integer, which is the finite-size effect we have seen in the main text, i.e., Fig.~\ref{fig:Z2ChernT3}, Fig.~\ref{fig:Z2ChernTd}, Fig.~\ref{fig:CKMH1_Z2Chern} and Fig.~\ref{fig:CKMH2_Z2Chern}, but we have also seen that the summation results converge to the quantized integer with increasing system size. 
Combining Eq.~(\ref{eq:CSImple1}) and Eq.~(\ref{eq:CSImple2}), we deduce the constant coefficient for Eq.~(\ref{eq:SquareProject}) as
\begin{eqnarray}
\label{eq:CSImple3}
\frac{1}{2\pi i}\cdot\frac{1}{4\delta_{q_u}\delta_{q_v}}\cdot\frac{\Omega_{BZ}}{N}=\frac{1}{2\pi i}\cdot\frac{3L_1L_2a^2}{16\pi^2}\cdot\frac{4\pi^2}{3L_1L_2a^2}=\frac{1}{8\pi i}.
\end{eqnarray}
So finally, we can get the simplified expression of Eq.~(\ref{eq:SquareProject}) as
\begin{eqnarray}
\label{eq:FinalChernNumber}
\mathcal{C}=&&\frac{1}{8\pi i}\sum_{i=1}^{L_1}\sum_{j=1}^{L_2}S(q_{u,i},q_{v,j}) \nonumber \\
S(q_{u,i},q_{v,j})=\text{Tr}\Big\{ Q_{i,j}\Big( [Q_{i+1,j},Q_{i,j+1}]&&+[Q_{i,j+1},Q_{i-1,j}]+[Q_{i-1,j},Q_{i,j-1}]+[Q_{i,j-1},Q_{i+1,j}] \Big) \Big\}.
\end{eqnarray}
During the calculation, we only need to  prepare the projection matrix $P(q_u,(-q_u+2q_v)/\sqrt{3})$ through the zero-frequency Green's function matrix obtained from QMC simulation. 
As Chern number should be an integer for a gapped system, we can only calculate the imaginary part of $S(q_{u,i},q_{v,j})$ in Eq.~(\ref{eq:FinalChernNumber}).


\section{Validity of Eq.~(\ref{eq:NewFourier1}) at zero-temperature}
\label{sec:appendix_b}
In this appendix, we validate the usage of Eq.~(\ref{eq:NewFourier1}) in obtaining $G(i\omega,\mathbf{k})$ data from $G(\tau,\mathbf{k})$ at zero temperature.

Let's start with a review on some basic facts about Matsubara frequency Green's function at finite-temperature. The Matsubara frequencies, $i\omega_{n}=i(2n+1)\pi/\beta,n\in Z$ for fermion systems and $i\omega_{n}=i2n\pi/\beta,n\in Z$ for boson systems, are actually the poles of the corresponding Fermi-Dirac and Bose-Einstein distribution function as
\begin{eqnarray}
\label{eq:Poles}
&& n_{FD}(\varepsilon)=\frac{1}{e^{\beta\varepsilon}+1}=\frac{1}{2}+\frac{1}{\beta}\sum_{n=-\infty}^{+\infty}\frac{1}{i\omega_n-\varepsilon} \hspace{1.0cm} \omega_n=\frac{(2n+1)\pi}{\beta}
\\ \nonumber
&& n_{BE}(\varepsilon)=\frac{1}{e^{\beta\varepsilon}-1}=-\frac{1}{2}-\frac{1}{\beta}\sum_{n=-\infty}^{+\infty}\frac{1}{i\omega_n-\varepsilon} \hspace{1.0cm} \omega_n=\frac{2n\pi}{\beta}.
\end{eqnarray}
We can explicitly observe that the exact zero Matsubara Frequency $i\omega=0$ can only be physically reachable at zero temperature. Otherwise, the imaginary-time single-particle Green's function at finite temperature is defined as
\begin{eqnarray}
\label{eq:MatsubaraGreenDef}
G(\tau,\hat{A}\hat{B}) = -\Big\langle T_{\tau}[\hat{A}(\tau)\hat{B}(0)] \Big\rangle = -\theta(\tau)\Big \langle e^{\tau\hat{H}}\hat{A}e^{-\tau\hat{H}}\hat{B} \Big\rangle \pm \theta(-\tau)\Big \langle \hat{B}e^{\tau\hat{H}}\hat{A}e^{-\tau\hat{H}} \Big\rangle,
\end{eqnarray}
where $+$ is for fermionic operators and $-$ is for bosonic operators, while $\hat{A}$ and $\hat{B}$ stands for single-particle fermionic or bosonic operators. According to this definition, one can simply prove the periodic and anti-periodic properties of $G(\tau,\hat{A}\hat{B})$ for fermionic and bosonic systems as
\begin{eqnarray}
\label{eq:Periodic}
{\text{Fermion}}&&: G(\tau,\hat{A}\hat{B})=-G(\beta+\tau,\hat{A}\hat{B}) \\ \nonumber
{\text{Boson}}&&:G(\tau,\hat{A}\hat{B})=G(\beta+\tau,\hat{A}\hat{B}).
\end{eqnarray}
We can also write down the Lehmann representation of $G(\tau,\hat{A}\hat{B})$ for both fermionic and bosonic systems via expanding the expectation by all many-body eigenstates of the system as
\begin{eqnarray}
\label{eq:LehmannGTau}
G(\tau,\hat{A}\hat{B}) = \frac{1}{Z}\sum_{mn}e^{(E_m-E_n)\tau}\Big[-\theta(\tau)e^{-\beta E_m}\pm\theta(-\tau)e^{-\beta E_n}\Big]
\langle m|\hat{A} |n\rangle \langle n|\hat{B} |m\rangle.
\end{eqnarray}
For finite temperature case, the Fourier transformation between the imaginary-time Green's function $G(\tau,\hat{A}\hat{B})$ and the Matsubara frequency Green's function $G(i\omega_n,\hat{A}\hat{B})$
\begin{eqnarray}
\label{eq:Transform}
G(i\omega_n,\hat{A}\hat{B})=\int_{0}^{\beta}G(\tau,\hat{A}\hat{B})e^{i\omega_n\tau}d\tau.
\end{eqnarray}
So the Lehmann representation expression of $G(i\omega_n,\hat{A}\hat{B})$ is
\begin{eqnarray}
\label{eq:FTLehmannGOmega}
G(i\omega_n,\hat{A}\hat{B}) = \sum_{mn}D_{mn}\frac{\langle n|\hat{A} |m\rangle \langle m|\hat{B} |n\rangle}{i\omega_n-(E_m-E_n)} \hspace{1.2cm} D_{mn}=\frac{\pm e^{-\beta E_m}+e^{-\beta E_n}}{Z},
\end{eqnarray}
where the sign $\pm$ originates from the term $e^{i\omega_n\beta}=+1$ for bosons and $e^{i\omega_n\beta}=-1$ for fermions. The above formulae summarize the basic properties of Matsubara frequency Green's function.

At exact zero-temperature, we have the Lehmann representation for $G(\tau,\mathbf{k})$ as
\begin{eqnarray}
\label{eq:ZeroTLehmannGTau}
G(\tau,\hat{A}\hat{B}) = \sum_{m}\Big[-\theta(\tau)e^{(E_0-E_m)\tau}\langle 0|\hat{A} |m\rangle \langle m|\hat{B} |0\rangle\pm\theta(-\tau)e^{-(E_0-E_m)\tau}\langle m|\hat{A} |0\rangle \langle 0|\hat{B} |m\rangle\Big].
\end{eqnarray}
Taking $\beta\to+\infty$ limit in Eq.~(\ref{eq:FTLehmannGOmega}), the Lehmann representation for $G(i\omega,\hat{A}\hat{B})$ at zero-temperature can also be reached as
\begin{eqnarray}
\label{eq:ZTTLehmannGOmega}
G(i\omega,\hat{A}\hat{B})=\sum_{m} \Big[\frac{\langle 0|\hat{A} |m\rangle \langle m|\hat{B} |0\rangle}{i\omega-(E_m-E_0)} \pm
\frac{\langle m|\hat{A} |0\rangle \langle 0|\hat{B} |m\rangle}{i\omega+(E_m-E_0)} \Big].
\end{eqnarray}
Now the problem is, at zero temperature, Fourier transformation in Eq.~(\ref{eq:Transform}) cannot transfer $G(\tau,\hat{A}\hat{B})$ in Eq.~(\ref{eq:ZeroTLehmannGTau}) to $G(i\omega,\hat{A}\hat{B})$ in Eq.~(\ref{eq:ZTTLehmannGOmega}), since Eq.~(\ref{eq:ZTTLehmannGOmega}) cannot be obtained if one naively takes $\beta\to+\infty$ limit in Eq.~(\ref{eq:Transform}), which corresponds to integration domain $[0,+\infty]$. To solve this problem, we present a simple generalization of finite temperature Matsubara frequency Green's function formalism into the zero-temperature case, by altering the Fourier transformation in Eq.~(\ref{eq:Transform}) to the following one as
\begin{eqnarray}
\label{eq:ZeroTransform}
G(i\omega,\hat{A}\hat{B})=\int_{-\infty}^{+\infty}G(\tau,\hat{A}\hat{B})e^{i\omega\tau}d\tau.
\end{eqnarray}
With this new transformation, we can now obtain Eq.~(\ref{eq:ZTTLehmannGOmega}) from Eq.~(\ref{eq:ZeroTLehmannGTau}), at zero-temperature. In practical QMC simulations, we do not take the integral in Eq.~(\ref{eq:ZeroTransform}) to $\pm\infty$, and instead we carry out a cutoff for $\tau$ in the integral as Eq.~(\ref{eq:NewFourier3}) in the main text. Nevertheless, Eq.~(\ref{eq:ZeroTransform}) validates the usage of Eq.~(\ref{eq:NewFourier2}) in the main text calculating $G(i\omega=0,\mathbf{k})$ from $G(\tau,\mathbf{k})$ matrix obtained from QMC simulations.

\section{Proof of Eq.~(\ref{eq:Relation1}), Eq.~(\ref{eq:Relation2}) and Eq.~(\ref{eq:Relation3})}
\label{sec:appendix_c}

The definition of $[G_{\sigma}(\tau,\mathbf{k})]_{pq}$ with $p,q=1,2,\cdots,m$ as orbitals for each spin sector in a unit cell, is
\begin{eqnarray}
\label{eq:E35}
[G_{\sigma}(\tau,\mathbf{k})]_{pq} = -\Big\langle T_{\tau}[c_{\mathbf{k}p{\sigma}}(\tau)c_{\mathbf{k}q{\sigma}}^\dagger(0)]\Big\rangle
=-\frac{1}{N}\sum_{i,j=1}^N e^{-i\mathbf{k}\cdot(\mathbf{R}_i-\mathbf{R}_j)}\Big\langle T_{\tau}[c_{ip{\sigma}}(\tau)c_{jq{\sigma}}^\dagger(0)]\Big\rangle ,
\end{eqnarray}
where $N=L^2$ is the number of unit cells. Then we have
\begin{eqnarray}
[G_{\sigma}(\tau,\mathbf{k})]_{qp}^\star
=-\frac{1}{N}\sum_{i,j=1}^N e^{i\mathbf{k}\cdot(\mathbf{R}_i-\mathbf{R}_j)}\Big\langle T_{\tau}[c_{iq{\sigma}}(\tau)c_{jp{\sigma}}^\dagger(0)]\Big\rangle^\star.
\end{eqnarray}
For any operator, we have $\langle\phi|\hat{A}|\phi\rangle^\star=\langle\phi|\hat{A}^\dagger|\phi\rangle$, which means $\langle\hat{A}\rangle^\star=\langle\hat{A}^\dagger\rangle$. Based on this relation, we have
\begin{eqnarray}
\Big\langle T_{\tau}[c_{iq{\sigma}}(\tau)c_{jp{\sigma}}^\dagger(0)]\Big\rangle^\star = \Big\langle T_{\tau}[c_{jp{\sigma}}(\tau)c_{iq{\sigma}}^\dagger(0)]\Big\rangle.
\end{eqnarray}
Then we can prove
\begin{eqnarray}
[G_{\sigma}(\tau,\mathbf{k})]_{qp}^\star
=&& -\frac{1}{N}\sum_{i,j=1}^N e^{i\mathbf{k}\cdot(\mathbf{R}_i-\mathbf{R}_j)}\Big\langle T_{\tau}[c_{jp{\sigma}}(\tau)c_{iq{\sigma}}^\dagger(0)]\Big\rangle
=-\frac{1}{N}\sum_{i,j=1}^N e^{-i\mathbf{k}\cdot(\mathbf{R}_i-\mathbf{R}_j)}\Big\langle T_{\tau}[c_{ip{\sigma}}(\tau)c_{jq{\sigma}}^\dagger(0)]\Big\rangle \nonumber \\
=&& [G_{\sigma}(\tau,\mathbf{k})]_{pq}. \hspace{0.5cm}
\end{eqnarray}
So the proof for Eq.~(\ref{eq:Relation1}) is complete.

For Eq.~(\ref{eq:Relation2}), the inversion operation $\hat{\mathcal{I}}$ (in each spin sector) transforms $p$ sublattice to $p^\prime$. For GKMH model, $\hat{\mathcal{I}}$ transfers A sublattice into B as $1\leftrightarrow2$, shown in Fig.~\ref{fig:KMHLatt} (a), while it transfers $1\leftrightarrow4$, $2\leftrightarrow5$ and $3\leftrightarrow6$ for CKMH model, shown in Fig.~\ref{fig:ClusterKMH}. We assume the relation $p\leftrightarrow p^\prime$ and $q\leftrightarrow q^\prime$ under the spatial inversion symmetry operation for generally multi-band systems. As the position vector should be inverse under $\hat{\mathcal{I}}$, i.e., $\mathbf{R}_{i}\to-\mathbf{R}_{i}$ we obtain the transformation for simple operators $c_{\mathbf{k}p\sigma},c_{\mathbf{k}q\sigma}$ in reciprocal space as
\begin{eqnarray}
&&c_{\mathbf{k}p\sigma}=\frac{1}{\sqrt{N}}\sum_{i}e^{-i\mathbf{k}\cdot\mathbf{R}_i}c_{ip\sigma}
\hspace{0.4cm} \to \nonumber
\\
&&\mathcal{I}c_{\mathbf{k}p\sigma}\mathcal{I}^{-1}=\frac{1}{\sqrt{N}}\sum_{i}\mathcal{I}e^{-i\mathbf{k}\cdot\mathbf{R}_i}\mathcal{I}^{-1}\cdot\mathcal{I}^{-1}c_{ip\sigma}\mathcal{I}^{-1}
=\frac{1}{\sqrt{N}}\sum_{i}e^{i\mathbf{k}\cdot\mathbf{R}_i}c_{ip^\prime\sigma}=c_{-\mathbf{k}p^\prime\sigma}.
\end{eqnarray}
Similarly, we have
\begin{eqnarray}
&& \mathcal{I}c_{\mathbf{k}p\sigma}^{\dagger}\mathcal{I}^{-1}=c_{-\mathbf{k}p^\prime\sigma}^{\dagger}
\nonumber\\
&& \mathcal{I}c_{\mathbf{k}q\sigma}(\tau)\mathcal{I}^{-1}=c_{-\mathbf{k}q^\prime\sigma}(\tau), \hspace{0.8cm} \mathcal{I}c_{\mathbf{k}q\sigma}^{\dagger}(\tau)\mathcal{I}^{-1}=c_{-\mathbf{k}q^\prime\sigma}^{\dagger}(\tau).
\end{eqnarray}
So for the imaginary-time Green's function matrix, we have
\begin{eqnarray}
\mathcal{I}[G_{\sigma}(\tau,\mathbf{k})]_{pq}\mathcal{I}^{-1}=&&-\Big\langle T_{\tau}[\mathcal{I}c_{\mathbf{k}p{\sigma}}(\tau)c_{\mathbf{k}q{\sigma}}^\dagger\mathcal{I}^{-1}]\Big\rangle
=-\Big\langle T_{\tau}[\mathcal{I}c_{\mathbf{k}p{\sigma}}(\tau)\mathcal{I}^{-1}\mathcal{I}c_{\mathbf{k}q{\sigma}}^\dagger\mathcal{I}^{-1}]\Big\rangle
=-\Big\langle c_{-\mathbf{k}p^\prime{\sigma}}(\tau)c_{-\mathbf{k}q^\prime{\sigma}}^\dagger\Big\rangle\\ \nonumber
=&& [G_{\sigma}(\tau,-\mathbf{k})]_{p^\prime q^\prime}
\end{eqnarray}
Due to the spatial inversion symmetry, the $G(\tau,\mathbf{k})$ matrix should be invariant under the inversion symmetry operation, from which we can get
\begin{eqnarray}
[G_{\sigma}(\tau,\mathbf{k})]_{pq}=[G_{\sigma}(\tau,-\mathbf{k})]_{p^\prime q^\prime}.
\end{eqnarray}
This is exactly the Eq.~(\ref{eq:Relation2}).

Finally, we prove the relation Eq.~(\ref{eq:Relation3}). 
The standard definition for $[G_{\sigma}(\tau,\mathbf{k})]_{pq}$ is
\begin{eqnarray}
\label{eq:Definition01}
[G_{\sigma}(\tau,\mathbf{k})]_{pq} &=& -\Big\langle T_{\tau}[c_{\mathbf{k}p{\sigma}}(\tau)c_{\mathbf{k}q{\sigma}}^\dagger(0)] \Big\rangle
=-\theta(\tau)\Big\langle c_{\mathbf{k}p{\sigma}}(\tau)c_{\mathbf{k}q{\sigma}}^\dagger \Big\rangle + \theta(-\tau)\Big\langle c_{\mathbf{k}q{\sigma}}^\dagger c_{\mathbf{k}p{\sigma}}(\tau) \Big\rangle
\\ \nonumber
&=& -\theta(\tau)\Big\langle c_{\mathbf{k}p{\sigma}}c_{\mathbf{k}q{\sigma}}^\dagger(-\tau) \Big\rangle + \theta(-\tau)\Big\langle c_{\mathbf{k}q{\sigma}}^\dagger (-\tau) c_{\mathbf{k}p{\sigma}} \Big\rangle
\\ \nonumber
&=& -\theta(\tau)\frac{1}{N}\sum_{i,j=1}^N e^{-i\mathbf{k}\cdot(\vec{R}_i-\vec{R}_j)}\Big\langle c_{ip{\sigma}}c_{jq{\sigma}}^\dagger(-\tau)\Big\rangle
+\theta(-\tau)\frac{1}{N}\sum_{i,j=1}^N e^{-i\mathbf{k}\cdot(\vec{R}_i-\vec{R}_j)}\Big\langle c_{jq{\sigma}}^\dagger(-\tau)c_{ip{\sigma}}\Big\rangle,
\end{eqnarray}
where we have applied the the following relations
\begin{eqnarray}
\label{eq:Relation01}
\Big\langle c_{\mathbf{k}p{\sigma}}(\tau)c_{\mathbf{k}q{\sigma}}^\dagger \Big\rangle = \Big\langle c_{\mathbf{k}p{\sigma}}c_{\mathbf{k}q{\sigma}}^\dagger(-\tau) \Big\rangle
\hspace{1.0cm}
\Big\langle c_{\mathbf{k}q{\sigma}}^\dagger c_{\mathbf{k}p{\sigma}}(\tau) \Big\rangle = \Big\langle c_{\mathbf{k}q{\sigma}}^\dagger (-\tau) c_{\mathbf{k}p{\sigma}} \Big\rangle.
\end{eqnarray}
Then we carry out the particle-hole transformation for $[G_{\sigma}(\tau,\mathbf{k})]_{pq}$ as $c_{p}\to\xi_pd_{p}^{\dagger}$ and $c_{q}\to\xi_qd_{q}^{\dagger}$, we have
\begin{eqnarray}
\frac{1}{N}\sum_{i,j=1}^N e^{-i\mathbf{k}\cdot(\mathbf{R}_i-\mathbf{R}_j)}\Big\langle c_{ip{\sigma}}c_{jq{\sigma}}^\dagger(-\tau)\Big\rangle
\hspace{0.3cm}\longrightarrow \hspace{0.3cm}&& \frac{1}{N}\sum_{i,j=1}^N e^{-i\mathbf{k}\cdot(\mathbf{R}_i-\mathbf{R}_j)}\xi_p\xi_q\Big\langle d_{ip{\sigma}}^{\dagger}d_{jq{\sigma}}(-\tau)\Big\rangle
\\ \nonumber
&&=\xi_p\xi_q\Big\langle d_{-\mathbf{k}p{\sigma}}^{\dagger}d_{-\mathbf{k}q{\sigma}}(-\tau)\Big\rangle,
\end{eqnarray}
and the other term
\begin{eqnarray}
\frac{1}{N}\sum_{i,j=1}^N e^{-i\mathbf{k}\cdot(\mathbf{R}_i-\mathbf{R}_j)}\Big\langle c_{jq{\sigma}}^\dagger(-\tau)c_{ip{\sigma}}\Big\rangle
\hspace{0.3cm}\longrightarrow \hspace{0.3cm}&& \frac{1}{N}\sum_{i,j=1}^N e^{-i\mathbf{k}\cdot(\mathbf{R}_i-\mathbf{R}_j)}\xi_p\xi_q\Big\langle d_{jq{\sigma}}(-\tau)d_{ip{\sigma}}^{\dagger}\Big\rangle
\\ \nonumber
&&=\xi_p\xi_q\Big\langle d_{-\mathbf{k}q{\sigma}}(-\tau)d_{-\mathbf{k}p{\sigma}}^{\dagger}\Big\rangle.
\end{eqnarray}
Combining the results of these two terms, under the particle-hole transformation, $[G_{\sigma}(\tau,\mathbf{k})]_{pq}$ changes to
\begin{eqnarray}
\label{eq:Relation11}
[G_{\sigma}(\tau,\mathbf{k})]_{pq} \hspace{0.3cm}\longrightarrow \hspace{0.3cm} -\theta(\tau)\xi_p\xi_q\Big\langle d_{-\mathbf{k}p{\sigma}}^{\dagger}d_{-\mathbf{k}q{\sigma}}(-\tau)\Big\rangle
+ \theta(-\tau)\xi_p\xi_q\Big\langle d_{-\mathbf{k}q{\sigma}}(-\tau)d_{-\mathbf{k}p{\sigma}}^{\dagger}\Big\rangle.
\end{eqnarray}
On the other hand, we can write down $[G_{\sigma}(-\tau,-\mathbf{k})]_{qp}$ according to original definition in Eq.~(\ref{eq:Definition01})
\begin{eqnarray}
\label{eq:Relation21}
[G_{\sigma}(-\tau,-\mathbf{k})]_{qp} = -\theta(-\tau)\Big\langle c_{-\mathbf{k}q{\sigma}}(-\tau)c_{-\mathbf{k}p{\sigma}}^{\dagger}\Big\rangle +\theta(\tau)\Big\langle c_{-\mathbf{k}p{\sigma}}^{\dagger}c_{-\mathbf{k}q{\sigma}}(-\tau)\Big\rangle
\end{eqnarray}
Comparing Eq.~(\ref{eq:Relation11}) and Eq.~(\ref{eq:Relation21}), and considering that the particle-hole symmetry is preserved, we arrive at
\begin{eqnarray}
\label{eq:Conclusion01}
[G_{\sigma}(\tau,\mathbf{k})]_{pq} = -\xi_p\xi_q[G_{\sigma}(-\tau,-\mathbf{k})]_{qp},
\end{eqnarray}
which is Eq.~(\ref{eq:Relation3}) in the main text.

\section{Derivation of Eq.~(\ref{eq:J22Equation}) and Eq.~(\ref{eq:J42Equation})}
\label{sec:appendix_d}

For the CKMH model, the spatial inversion operators as $1\leftrightarrow4$, $2\leftrightarrow5$ and $3\leftrightarrow6$ among the six sublattices. One can  show that spatial inversion symmetry result in the following properties of $G_{\sigma}(\tau,\mathbf{k})$ matrix, according to Eq.~(\ref{eq:Relation2}).
\begin{eqnarray}
\label{eq:J5Equation}
&& [G_{\sigma}(\tau,\mathbf{k})]_{11}=[G_{\sigma}(\tau,-\mathbf{k})]_{44}
   \hspace{1.0cm} [G_{\sigma}(\tau,\mathbf{k})]_{12}=[G_{\sigma}(\tau,-\mathbf{k})]_{45}
   \hspace{1.0cm} [G_{\sigma}(\tau,\mathbf{k})]_{13}=[G_{\sigma}(\tau,-\mathbf{k})]_{46}
\nonumber \\
&& [G_{\sigma}(\tau,\mathbf{k})]_{14}=[G_{\sigma}(\tau,-\mathbf{k})]_{41}
   \hspace{1.0cm} [G_{\sigma}(\tau,\mathbf{k})]_{15}=[G_{\sigma}(\tau,-\mathbf{k})]_{42}
   \hspace{1.0cm} [G_{\sigma}(\tau,\mathbf{k})]_{16}=[G_{\sigma}(\tau,-\mathbf{k})]_{43}
\nonumber \\
&& [G_{\sigma}(\tau,\mathbf{k})]_{22}=[G_{\sigma}(\tau,-\mathbf{k})]_{55}
   \hspace{1.0cm} [G_{\sigma}(\tau,\mathbf{k})]_{23}=[G_{\sigma}(\tau,-\mathbf{k})]_{56}
   \hspace{1.0cm} [G_{\sigma}(\tau,\mathbf{k})]_{24}=[G_{\sigma}(\tau,-\mathbf{k})]_{51}
\nonumber \\
&& [G_{\sigma}(\tau,\mathbf{k})]_{25}=[G_{\sigma}(\tau,-\mathbf{k})]_{52}
   \hspace{1.0cm} [G_{\sigma}(\tau,\mathbf{k})]_{26}=[G_{\sigma}(\tau,-\mathbf{k})]_{53}
\nonumber \\
&& [G_{\sigma}(\tau,\mathbf{k})]_{33}=[G_{\sigma}(\tau,-\mathbf{k})]_{66}
   \hspace{1.0cm} [G_{\sigma}(\tau,\mathbf{k})]_{34}=[G_{\sigma}(\tau,-\mathbf{k})]_{61}
   \hspace{1.0cm} [G_{\sigma}(\tau,\mathbf{k})]_{35}=[G_{\sigma}(\tau,-\mathbf{k})]_{62}
\nonumber \\
&& [G_{\sigma}(\tau,\mathbf{k})]_{36}=[G_{\sigma}(\tau,-\mathbf{k})]_{63}
\nonumber \\
&& [G_{\sigma}(\tau,\mathbf{k})]_{44}=[G_{\sigma}(\tau,-\mathbf{k})]_{11}
   \hspace{1.0cm} [G_{\sigma}(\tau,\mathbf{k})]_{45}=[G_{\sigma}(\tau,-\mathbf{k})]_{12}
   \hspace{1.0cm} [G_{\sigma}(\tau,\mathbf{k})]_{46}=[G_{\sigma}(\tau,-\mathbf{k})]_{13}
\nonumber \\
&& [G_{\sigma}(\tau,\mathbf{k})]_{55}=[G_{\sigma}(\tau,-\mathbf{k})]_{22}
   \hspace{1.0cm} [G_{\sigma}(\tau,\mathbf{k})]_{56}=[G_{\sigma}(\tau,-\mathbf{k})]_{23}
\nonumber \\
&& [G_{\sigma}(\tau,\mathbf{k})]_{66}=[G_{\sigma}(\tau,-\mathbf{k})]_{33}
\end{eqnarray}
The particle-hole operators with $\xi_1=+1,\xi_2=-1,\xi_3=+1,\xi_4=-1,\xi_5=+1,\xi_6=-1$, then according to Eq.~(\ref{eq:Relation3}), we have the following properties for $[G_{\sigma}(\tau,\mathbf{k})]_{1q},q=1,2,3,4,5,6$ as
\begin{eqnarray}
\label{eq:J7Equation}
&& [G_{\sigma}(\tau,-\mathbf{k})]_{44}=-[G_{\sigma}(-\tau,\mathbf{k})]_{44}
   \hspace{1.0cm} [G_{\sigma}(\tau,-\mathbf{k})]_{45}=+[G_{\sigma}(-\tau,\mathbf{k})]_{54}
   \hspace{1.0cm} [G_{\sigma}(\tau,-\mathbf{k})]_{46}=-[G_{\sigma}(-\tau,\mathbf{k})]_{64}  \hspace{0.5cm}
\nonumber \\
&& [G_{\sigma}(\tau,-\mathbf{k})]_{41}=+[G_{\sigma}(-\tau,\mathbf{k})]_{14}
   \hspace{1.0cm} [G_{\sigma}(\tau,-\mathbf{k})]_{42}=-[G_{\sigma}(-\tau,\mathbf{k})]_{24}
   \hspace{1.0cm} [G_{\sigma}(\tau,-\mathbf{k})]_{43}=+[G_{\sigma}(-\tau,\mathbf{k})]_{34}.
\end{eqnarray}
With the condition of Hermitian matrix of $G_{\sigma}(\tau,\mathbf{k})$ in Eq.~(\ref{eq:Relation1}), we have $[G_{\sigma}(-\tau,\mathbf{k})]_{54}=[G_{\sigma}(-\tau,\mathbf{k})]_{45}^{\star}, [G_{\sigma}(-\tau,\mathbf{k})]_{64}=[G_{\sigma}(-\tau,\mathbf{k})]_{46}^{\star}$. With Eq.~(\ref{eq:J5Equation}) and Eq.~(\ref{eq:J7Equation}), we have
\begin{eqnarray}
\label{eq:J8Equation}
&& [G_{\sigma}(\tau,\mathbf{k})]_{11}=-[G_{\sigma}(-\tau,\mathbf{k})]_{44}
   \hspace{1.0cm} [G_{\sigma}(\tau,\mathbf{k})]_{12}=+[G_{\sigma}(-\tau,\mathbf{k})]_{45}^{\star}
   \hspace{1.0cm} [G_{\sigma}(\tau,\mathbf{k})]_{13}=-[G_{\sigma}(-\tau,\mathbf{k})]_{46}^{\star}
\nonumber \\
&& [G_{\sigma}(\tau,\mathbf{k})]_{14}=+[G_{\sigma}(-\tau,\mathbf{k})]_{14}
   \hspace{1.0cm} [G_{\sigma}(\tau,\mathbf{k})]_{15}=-[G_{\sigma}(-\tau,\mathbf{k})]_{24}
   \hspace{1.0cm} [G_{\sigma}(\tau,\mathbf{k})]_{16}=+[G_{\sigma}(-\tau,\mathbf{k})]_{34}.
\end{eqnarray}
Similarly, for $[G_{\sigma}(\tau,\mathbf{k})]_{2q},q=2,3,4,5,6$ from the particle-hole symmetry, we have
\begin{eqnarray}
\label{eq:J9Equation}
&& [G_{\sigma}(\tau,-\mathbf{k})]_{55}=-[G_{\sigma}(-\tau,\mathbf{k})]_{55}
   \hspace{1.0cm} [G_{\sigma}(\tau,-\mathbf{k})]_{56}=+[G_{\sigma}(-\tau,\mathbf{k})]_{65}
   \hspace{1.0cm} [G_{\sigma}(\tau,-\mathbf{k})]_{51}=-[G_{\sigma}(-\tau,\mathbf{k})]_{15} \hspace{0.5cm}
\nonumber \\
&& [G_{\sigma}(\tau,-\mathbf{k})]_{52}=+[G_{\sigma}(-\tau,\mathbf{k})]_{25}
   \hspace{1.0cm} [G_{\sigma}(\tau,-\mathbf{k})]_{53}=-[G_{\sigma}(-\tau,\mathbf{k})]_{35}.
\end{eqnarray}
With $[G_{\sigma}(-\tau,\mathbf{k})]_{65}=[G_{\sigma}(-\tau,\mathbf{k})]_{56}^{\star}$ and combining Eq.~(\ref{eq:J5Equation}) and Eq.~(\ref{eq:J9Equation}), we also have
\begin{eqnarray}
\label{eq:J10Equation}
&& [G_{\sigma}(\tau,\mathbf{k})]_{22}=-[G_{\sigma}(-\tau,\mathbf{k})]_{55}
   \hspace{1.0cm} [G_{\sigma}(\tau,\mathbf{k})]_{23}=+[G_{\sigma}(-\tau,\mathbf{k})]_{56}^{\star}
   \hspace{1.0cm} [G_{\sigma}(\tau,\mathbf{k})]_{24}=-[G_{\sigma}(-\tau,\mathbf{k})]_{15}
\nonumber \\
&& [G_{\sigma}(\tau,\mathbf{k})]_{25}=+[G_{\sigma}(-\tau,\mathbf{k})]_{25}
   \hspace{1.0cm} [G_{\sigma}(\tau,\mathbf{k})]_{26}=-[G_{\sigma}(-\tau,\mathbf{k})]_{35}.
\end{eqnarray}
For $[G_{\sigma}(\tau,\mathbf{k})]_{3q},q=3,4,5,6$ from the particle-hole symmetry, we have
\begin{eqnarray}
\label{eq:J11Equation}
&& [G_{\sigma}(\tau,-\mathbf{k})]_{66}=-[G_{\sigma}(-\tau,\mathbf{k})]_{66}
   \hspace{1.0cm} [G_{\sigma}(\tau,-\mathbf{k})]_{61}=+[G_{\sigma}(-\tau,\mathbf{k})]_{16}
   \hspace{1.0cm} [G_{\sigma}(\tau,-\mathbf{k})]_{62}=-[G_{\sigma}(-\tau,\mathbf{k})]_{26} \hspace{0.5cm}
\nonumber \\
&& [G_{\sigma}(\tau,-\mathbf{k})]_{63}=+[G_{\sigma}(-\tau,\mathbf{k})]_{36}.
\end{eqnarray}
Combining Eq.~(\ref{eq:J5Equation}) and Eq.~(\ref{eq:J11Equation}), we also have
\begin{eqnarray}
\label{eq:J12Equation}
&& [G_{\sigma}(\tau,\mathbf{k})]_{33}=-[G_{\sigma}(-\tau,\mathbf{k})]_{66}
   \hspace{1.0cm} [G_{\sigma}(\tau,\mathbf{k})]_{34}=+[G_{\sigma}(-\tau,\mathbf{k})]_{16}
   \hspace{1.0cm} [G_{\sigma}(\tau,\mathbf{k})]_{35}=-[G_{\sigma}(-\tau,\mathbf{k})]_{26} \hspace{0.5cm}
\nonumber \\
&& [G_{\sigma}(\tau,\mathbf{k})]_{36}=+[G_{\sigma}(-\tau,\mathbf{k})]_{36}.
\end{eqnarray}
For $[G_{\sigma}(\tau,\mathbf{k})]_{4q},q=4,5,6$ from the particle-hole symmetry, we have
\begin{eqnarray}
\label{eq:J13Equation}
[G_{\sigma}(\tau,-\mathbf{k})]_{11} = -[G_{\sigma}(-\tau,\mathbf{k})]_{11}
\hspace{1.0cm} [G_{\sigma}(\tau,-\mathbf{k})]_{12}=+[G_{\sigma}(-\tau,\mathbf{k})]_{21}
\hspace{1.0cm} [G_{\sigma}(\tau,-\mathbf{k})]_{13}=-[G_{\sigma}(-\tau,\mathbf{k})]_{31} \hspace{0.5cm}
\end{eqnarray}
With $[G_{\sigma}(-\tau,\mathbf{k})]_{21}=[G_{\sigma}(-\tau,\mathbf{k})]_{12}^{\star}, [G_{\sigma}(-\tau,\mathbf{k})]_{31}=[G_{\sigma}(-\tau,\mathbf{k})]_{13}^{\star}$ and combining Eq.~(\ref{eq:J5Equation}) and Eq.~(\ref{eq:J13Equation}), we also have
\begin{eqnarray}
\label{eq:J14Equation}
 [G_{\sigma}(\tau,\mathbf{k})]_{44}=-[G_{\sigma}(-\tau,\mathbf{k})]_{11}
   \hspace{1.0cm} [G_{\sigma}(\tau,\mathbf{k})]_{45}=+[G_{\sigma}(-\tau,\mathbf{k})]_{12}^{\star}
   \hspace{1.0cm} [G_{\sigma}(\tau,\mathbf{k})]_{46}=-[G_{\sigma}(-\tau,\mathbf{k})]_{13}^{\star}
\end{eqnarray}
For $[G_{\sigma}(\tau,\mathbf{k})]_{5q},q=5,6$ from the particle-hole symmetry, we have
\begin{eqnarray}
\label{eq:J15Equation}
[G_{\sigma}(\tau,-\mathbf{k})]_{22}=-[G_{\sigma}(-\tau,\mathbf{k})]_{22} \hspace{1.0cm}
[G_{\sigma}(\tau,-\mathbf{k})]_{23}=+[G_{\sigma}(-\tau,\mathbf{k})]_{32}
\end{eqnarray}
With $[G_{\sigma}(-\tau,\mathbf{k})]_{32}=[G_{\sigma}(-\tau,\mathbf{k})]_{23}^{\star}$ and combining Eq.~(\ref{eq:J5Equation}) and Eq.~(\ref{eq:J15Equation}), we also have
\begin{eqnarray}
\label{eq:J16Equation}
[G_{\sigma}(\tau,\mathbf{k})]_{55}=-[G_{\sigma}(-\tau,\mathbf{k})]_{22}
\hspace{1.0cm} [G_{\sigma}(\tau,\mathbf{k})]_{56}=+[G_{\sigma}(-\tau,\mathbf{k})]_{23}^{\star}
\end{eqnarray}
For $[G_{\sigma}(\tau,\mathbf{k})]_{66}$ from the particle-hole symmetry, we have
\begin{eqnarray}
\label{eq:J17Equation}
[G_{\sigma}(\tau,-\mathbf{k})]_{33}=-[G_{\sigma}(-\tau,\mathbf{k})]_{33}
\end{eqnarray}
Combining Eq.~(\ref{eq:J5Equation}) and Eq.~(\ref{eq:J17Equation}), we also have
\begin{eqnarray}
\label{eq:J18Equation}
[G_{\sigma}(\tau,\mathbf{k})]_{66}=-[G_{\sigma}(-\tau,\mathbf{k})]_{33}
\end{eqnarray}
To calculate the $G_{\sigma}(i\omega=0,\mathbf{k})$ matrix, we need to apply the Fourier transformation as
\begin{eqnarray}
\label{eq:J19Equation}
G_{\sigma}(i\omega=0,\mathbf{k}) = \int_{-\infty}^{+\infty}G_{\sigma}(\tau,\mathbf{k})d\tau.
\end{eqnarray}
Then by Eq.~(\ref{eq:J19Equation}), we can show that there are actually only 12 independent matrix elements in $G_{\sigma}(i\omega=0,\mathbf{k})$ matrix for CKMH model, which we need to calculate, combining Eq.~(\ref{eq:J8Equation}), Eq.~(\ref{eq:J10Equation}), Eq.~(\ref{eq:J12Equation}), Eq.~(\ref{eq:J14Equation}), Eq.~(\ref{eq:J16Equation}) and Eq.~(\ref{eq:J18Equation}). For the diagonal matrix elements of $G_{\sigma}(i\omega=0,\mathbf{k})$ matrix, we have 3 independent diagonal elements as
\begin{eqnarray}
\label{eq:J20Equation}
&& A_1=[G_{\sigma}(i\omega=0,\mathbf{k})]_{11} = \int_{-\infty}^{+\infty}[G_{\sigma}(\tau,\mathbf{k})]_{11}d\tau
                                        = \int_{0}^{+\infty}\Big\{[G_{\sigma}(\tau,\mathbf{k})]_{11}-[G_{\sigma}(\tau,\mathbf{k})]_{44}\Big\}d\tau
                                        = -[G_{\sigma}(i\omega=0,\mathbf{k})]_{44} \hspace{0.6cm}
\nonumber \\
&& A_2=[G_{\sigma}(i\omega=0,\mathbf{k})]_{22} = \int_{-\infty}^{+\infty}[G_{\sigma}(\tau,\mathbf{k})]_{22}d\tau
                                        = \int_{0}^{+\infty}\Big\{[G_{\sigma}(\tau,\mathbf{k})]_{22}-[G_{\sigma}(\tau,\mathbf{k})]_{55}\Big\}d\tau
                                        = -[G_{\sigma}(i\omega=0,\mathbf{k})]_{55}
\nonumber \\
&& A_3=[G_{\sigma}(i\omega=0,\mathbf{k})]_{33} = \int_{-\infty}^{+\infty}[G_{\sigma}(\tau,\mathbf{k})]_{33}d\tau
                                        = \int_{0}^{+\infty}\Big\{[G_{\sigma}(\tau,\mathbf{k})]_{33}-[G_{\sigma}(\tau,\mathbf{k})]_{66}\Big\}d\tau
                                        = -[G_{\sigma}(i\omega=0,\mathbf{k})]_{66}
\end{eqnarray}
For the off-diagonal matrix elements, we can determine that there are only 9 independent matrix off-diagonal elements in $G_{\sigma}(i\omega=0,\mathbf{k})$ matrix as
\begin{eqnarray}
\label{eq:J21Equation}
&& A_4=[G_{\sigma}(i\omega=0,\mathbf{k})]_{12} = \int_{-\infty}^{+\infty}[G_{\sigma}(\tau,\mathbf{k})]_{12}d\tau
                                        = \int_{0}^{+\infty}\Big\{[G_{\sigma}(\tau,\mathbf{k})]_{12}+[G_{\sigma}(\tau,\mathbf{k})]_{45}^{\star}\Big\}d\tau
                                        = +[G_{\sigma}(i\omega=0,\mathbf{k})]_{45}^{\star} \hspace{0.5cm}
\nonumber \\
&& A_5=[G_{\sigma}(i\omega=0,\mathbf{k})]_{13} = \int_{-\infty}^{+\infty}[G_{\sigma}(\tau,\mathbf{k})]_{13}d\tau
                                        = \int_{0}^{+\infty}\Big\{[G_{\sigma}(\tau,\mathbf{k})]_{13}-[G_{\sigma}(\tau,\mathbf{k})]_{46}^{\star}\Big\}d\tau
                                        = -[G_{\sigma}(i\omega=0,\mathbf{k})]_{46}^{\star}
\nonumber \\
&& A_6=[G_{\sigma}(i\omega=0,\mathbf{k})]_{14} = \int_{-\infty}^{+\infty}[G_{\sigma}(\tau,\mathbf{k})]_{14}d\tau
                                        = 2\int_{0}^{+\infty}[G_{\sigma}(\tau,\mathbf{k})]_{14}d\tau
\nonumber \\
&& A_7=[G_{\sigma}(i\omega=0,\mathbf{k})]_{15} = \int_{-\infty}^{+\infty}[G_{\sigma}(\tau,\mathbf{k})]_{15}d\tau
                                        = \int_{0}^{+\infty}\Big\{[G_{\sigma}(\tau,\mathbf{k})]_{15}-[G_{\sigma}(\tau,\mathbf{k})]_{24}\Big\}d\tau
                                        = -[G_{\sigma}(i\omega=0,\mathbf{k})]_{24}
\nonumber \\
&& A_8=[G_{\sigma}(i\omega=0,\mathbf{k})]_{16} = \int_{-\infty}^{+\infty}[G_{\sigma}(\tau,\mathbf{k})]_{16}d\tau
                                        = \int_{0}^{+\infty}\Big\{[G_{\sigma}(\tau,\mathbf{k})]_{16}+[G_{\sigma}(\tau,\mathbf{k})]_{34}\Big\}d\tau
                                        = +[G_{\sigma}(i\omega=0,\mathbf{k})]_{34}
\nonumber \\
&& A_9=[G_{\sigma}(i\omega=0,\mathbf{k})]_{23} = \int_{-\infty}^{+\infty}[G_{\sigma}(\tau,\mathbf{k})]_{23}d\tau
                                        = \int_{0}^{+\infty}\Big\{[G_{\sigma}(\tau,\mathbf{k})]_{23}+[G_{\sigma}(\tau,\mathbf{k})]_{56}^{\star}\Big\}d\tau
                                        = +[G_{\sigma}(i\omega=0,\mathbf{k})]_{56}^{\star}
\nonumber \\
&& A_{10}=[G_{\sigma}(i\omega=0,\mathbf{k})]_{25} = \int_{-\infty}^{+\infty}[G_{\sigma}(\tau,\mathbf{k})]_{25}d\tau
                                        = 2\int_{0}^{+\infty}[G_{\sigma}(\tau,\mathbf{k})]_{25}d\tau
\nonumber \\
&& A_{11}=[G_{\sigma}(i\omega=0,\mathbf{k})]_{26} = \int_{-\infty}^{+\infty}[G_{\sigma}(\tau,\mathbf{k})]_{26}d\tau
                                        = \int_{0}^{+\infty}\Big\{[G_{\sigma}(\tau,\mathbf{k})]_{26}-[G_{\sigma}(\tau,\mathbf{k})]_{35}\Big\}d\tau
                                        = -[G_{\sigma}(i\omega=0,\mathbf{k})]_{35}
\nonumber \\
&& A_{12}=[G_{\sigma}(i\omega=0,\mathbf{k})]_{36} = \int_{-\infty}^{+\infty}[G_{\sigma}(\tau,\mathbf{k})]_{36}d\tau
                                        = 2\int_{0}^{+\infty}[G_{\sigma}(\tau,\mathbf{k})]_{36}d\tau
\end{eqnarray}
Then taking the $\tau$ cutoff $\theta$ and transforming the integral to summation over discrete $\tau$, we can get the following matrix structure of $G_{\sigma}(i\omega=0,\mathbf{k})$ matrix as
\begin{eqnarray}
G_{\sigma}(i\omega=0,\mathbf{k})=\left(                 
  \begin{array}{cccccc}   
    A_1         &   A_4           &   A_5            &   A_6   &   A_7           &   A_8            \\
    A_4^\star   &   A_2           &   A_9            &  -A_7   &   A_{10}        &   A_{11}         \\
    A_5^\star   &   A_9^\star     &   A_3            &   A_8   &  -A_{11}        &   A_{12}         \\
    A_6^\star   &  -A_7^\star     &   A_8^\star      &  -A_1   &   A_4^{\star}   &  -A_5^{\star}    \\
    A_7^\star   &   A_{10}^\star  &  -A_{11}^\star   &   A_4   &  -A_2           &   A_9^{\star}    \\
    A_8^\star   &   A_{11}^\star  &   A_{12}^\star   &  -A_5   &   A_9           &  -A_3
  \end{array}
    \right).
\end{eqnarray}

As for the $Z_2$ invariant for CKMH model, we only need to obtain the $G_{\sigma}(i\omega=0,\boldsymbol{\kappa})$ data at four TRIM points as $\boldsymbol{\kappa}=\boldsymbol{\Gamma},\mathbf{M}_1,\mathbf{M}_2,\mathbf{M}_3$. From the symmetry properties in Eq.~(\ref{eq:J5Equation}), we can obtain that for the TRIM points, we have
\begin{eqnarray}
\label{eq:J35Equation}
&& [G_{\sigma}(\tau,\boldsymbol{\kappa})]_{11}=[G_{\sigma}(\tau,\boldsymbol{\kappa})]_{44}
   \hspace{1.0cm} [G_{\sigma}(\tau,\boldsymbol{\kappa})]_{12}=[G_{\sigma}(\tau,\boldsymbol{\kappa})]_{45}
   \hspace{1.0cm} [G_{\sigma}(\tau,\boldsymbol{\kappa})]_{13}=[G_{\sigma}(\tau,\boldsymbol{\kappa})]_{46}
\nonumber \\
&& [G_{\sigma}(\tau,\boldsymbol{\kappa})]_{14}=[G_{\sigma}(\tau,\boldsymbol{\kappa})]_{41}
   \hspace{1.0cm} [G_{\sigma}(\tau,\boldsymbol{\kappa})]_{15}=[G_{\sigma}(\tau,\boldsymbol{\kappa})]_{42}
   \hspace{1.0cm} [G_{\sigma}(\tau,\boldsymbol{\kappa})]_{16}=[G_{\sigma}(\tau,\boldsymbol{\kappa})]_{43}
\nonumber \\
&& [G_{\sigma}(\tau,\boldsymbol{\kappa})]_{22}=[G_{\sigma}(\tau,\boldsymbol{\kappa})]_{55}
   \hspace{1.0cm} [G_{\sigma}(\tau,\boldsymbol{\kappa})]_{23}=[G_{\sigma}(\tau,\boldsymbol{\kappa})]_{56}
   \hspace{1.0cm} [G_{\sigma}(\tau,\boldsymbol{\kappa})]_{24}=[G_{\sigma}(\tau,\boldsymbol{\kappa})]_{51}
\nonumber \\
&& [G_{\sigma}(\tau,\boldsymbol{\kappa})]_{25}=[G_{\sigma}(\tau,\boldsymbol{\kappa})]_{52}
   \hspace{1.0cm} [G_{\sigma}(\tau,\boldsymbol{\kappa})]_{26}=[G_{\sigma}(\tau,\boldsymbol{\kappa})]_{53}
\nonumber \\
&& [G_{\sigma}(\tau,\boldsymbol{\kappa})]_{33}=[G_{\sigma}(\tau,\boldsymbol{\kappa})]_{66}
   \hspace{1.0cm} [G_{\sigma}(\tau,\boldsymbol{\kappa})]_{34}=[G_{\sigma}(\tau,\boldsymbol{\kappa})]_{61}
   \hspace{1.0cm} [G_{\sigma}(\tau,\boldsymbol{\kappa})]_{35}=[G_{\sigma}(\tau,\boldsymbol{\kappa})]_{62}
\nonumber \\
&& [G_{\sigma}(\tau,\boldsymbol{\kappa})]_{36}=[G_{\sigma}(\tau,\boldsymbol{\kappa})]_{63}
\nonumber \\
&& [G_{\sigma}(\tau,\boldsymbol{\kappa})]_{44}=[G_{\sigma}(\tau,\boldsymbol{\kappa})]_{11}
   \hspace{1.0cm} [G_{\sigma}(\tau,\boldsymbol{\kappa})]_{45}=[G_{\sigma}(\tau,\boldsymbol{\kappa})]_{12}
   \hspace{1.0cm} [G_{\sigma}(\tau,\boldsymbol{\kappa})]_{46}=[G_{\sigma}(\tau,\boldsymbol{\kappa})]_{13}
\nonumber \\
&& [G_{\sigma}(\tau,\boldsymbol{\kappa})]_{55}=[G_{\sigma}(\tau,\boldsymbol{\kappa})]_{22}
   \hspace{1.0cm} [G_{\sigma}(\tau,\boldsymbol{\kappa})]_{56}=[G_{\sigma}(\tau,\boldsymbol{\kappa})]_{23}
\nonumber \\
&& [G_{\sigma}(\tau,\boldsymbol{\kappa})]_{66}=[G_{\sigma}(\tau,\boldsymbol{\kappa})]_{33}
\end{eqnarray}
Then we can determine that at TRIM points, we have $A_1=A_2=A_3=0$ and $[G_{\sigma}(\tau,\boldsymbol{\kappa})]_{14}$, $[G_{\sigma}(\tau,\boldsymbol{\kappa})]_{25}$ and $[G_{\sigma}(\tau,\boldsymbol{\kappa})]_{36}$ are all real numbers. From these relations, we have $B_6=A_6$, $B_{10}=A_{10}$ and $B_{12}=A_{12}$, and they are all real numbers as well. For other elements, we have
\begin{eqnarray}
&& A_4=[G_{\sigma}(i\omega=0,\boldsymbol{\kappa})]_{12}=\int_{-\infty}^{+\infty}[G_{\sigma}(\tau,\boldsymbol{\kappa})]_{12}d\tau
      =\int_{0}^{+\infty}\Big\{[G_{\sigma}(\tau,\boldsymbol{\kappa})]_{12}+[G_{\sigma}(\tau,\boldsymbol{\kappa})]_{12}^{\star}\Big\}d\tau
      = B_4  \nonumber \\
&& A_5=[G_{\sigma}(i\omega=0,\boldsymbol{\kappa})]_{13} = \int_{-\infty}^{+\infty}[G_{\sigma}(\tau,\boldsymbol{\kappa})]_{13}d\tau
      = \int_{0}^{+\infty}\Big\{[G_{\sigma}(\tau,\boldsymbol{\kappa})]_{13}-[G_{\sigma}(\tau,\boldsymbol{\kappa})]_{13}^{\star}\Big\}d\tau
      = iB_5  \nonumber \\
&& A_6=[G_{\sigma}(i\omega=0,\boldsymbol{\kappa})]_{14} = \int_{-\infty}^{+\infty}[G_{\sigma}(\tau,\boldsymbol{\kappa})]_{14}d\tau
      = 2\int_{0}^{+\infty}[G_{\sigma}(\tau,\boldsymbol{\kappa})]_{14}d\tau
      = B_6  \nonumber \\
&& A_7=[G_{\sigma}(i\omega=0,\boldsymbol{\kappa})]_{15} = \int_{-\infty}^{+\infty}[G_{\sigma}(\tau,\boldsymbol{\kappa})]_{15}d\tau
      =\int_{0}^{+\infty}\Big\{[G_{\sigma}(\tau,\boldsymbol{\kappa})]_{15}-[G_{\sigma}(\tau,\boldsymbol{\kappa})]_{15}^{\star}\Big\}d\tau
      = iB_7  \nonumber \\
&& A_8=[G_{\sigma}(i\omega=0,\boldsymbol{\kappa})]_{16} = \int_{-\infty}^{+\infty}[G_{\sigma}(\tau,\boldsymbol{\kappa})]_{16}d\tau
      =\int_{0}^{+\infty}\Big\{[G_{\sigma}(\tau,\boldsymbol{\kappa})]_{16}+[G_{\sigma}(\tau,\boldsymbol{\kappa})]_{16}^{\star}\Big\}d\tau
      = B_8  \nonumber \\
&& A_9=[G_{\sigma}(i\omega=0,\boldsymbol{\kappa})]_{23} = \int_{-\infty}^{+\infty}[G_{\sigma}(\tau,\boldsymbol{\kappa})]_{23}d\tau
      = \int_{0}^{+\infty}\Big\{[G_{\sigma}(\tau,\boldsymbol{\kappa})]_{23}+[G_{\sigma}(\tau,\boldsymbol{\kappa})]_{23}^{\star}\Big\}d\tau
      = B_9  \nonumber \\
&& A_{10}=[G_{\sigma}(i\omega=0,\boldsymbol{\kappa})]_{25} = \int_{-\infty}^{+\infty}[G_{\sigma}(\tau,\boldsymbol{\kappa})]_{25}d\tau
         = 2\int_{0}^{+\infty}[G_{\sigma}(\tau,\boldsymbol{\kappa})]_{25}d\tau
         = B_{10}   \nonumber \\
&& A_{11}=[G_{\sigma}(i\omega=0,\boldsymbol{\kappa})]_{26} = \int_{-\infty}^{+\infty}[G_{\sigma}(\tau,\boldsymbol{\kappa})]_{26}d\tau
         = \int_{0}^{+\infty}\Big\{[G_{\sigma}(\tau,\boldsymbol{\kappa})]_{26}-[G_{\sigma}(\tau,\boldsymbol{\kappa})]_{26}^{\star}\Big\}d\tau
         = iB_{11}  \nonumber \\
&& A_{12}=[G_{\sigma}(i\omega=0,\mathbf{k})]_{36} = \int_{-\infty}^{+\infty}[G_{\sigma}(\tau,\mathbf{k})]_{36}d\tau
         = 2\int_{0}^{+\infty}[G_{\sigma}(\tau,\mathbf{k})]_{36}d\tau
         = B_{12}
\end{eqnarray}
We can observe that all the matrix elements of $G_{\sigma}(\tau,\boldsymbol{\kappa})$ matrix must be either purely real or purely imaginary, and the diagonal matrix elements are all zero. Hence, the matrix structure as $G_{\sigma}(\tau,\boldsymbol{\kappa})$ is as following,
\begin{eqnarray}
\label{eq:J234Equation}
G_{\sigma}(i\omega=0,\boldsymbol{\kappa})=\left(                 
  \begin{array}{cccccc}   
    0     &   B_4     &   iB_5     &   B_6   &   iB_7          &   B_8            \\
    B_4   &   0       &   B_9      &  -iB_7  &   B_{10}        &   iB_{11}         \\
   -iB_5  &   B_9     &   0        &   B_8   &  -iB_{11}       &   B_{12}         \\
    B_6   &  iB_7     &   B_8      &  0      &   B_4           &  iB_5    \\
   -iB_7  &   B_{10}  &  iB_{11}   &   B_4   &    0            &   B_9    \\
    B_8   & -iB_{11}  &   B_{12}   & -iB_5   &   B_9           &  0
  \end{array}
    \right),
\end{eqnarray}
in which $\{B_i|i=4,5,\cdots,12\}$ are purely real numbers. 

\twocolumngrid
\bibliography{SpinChernI_Bib}

\begin{thebibliography}{46}%
\makeatletter
\providecommand \@ifxundefined [1]{%
 \@ifx{#1\undefined}
}%
\providecommand \@ifnum [1]{%
 \ifnum #1\expandafter \@firstoftwo
 \else \expandafter \@secondoftwo
 \fi
}%
\providecommand \@ifx [1]{%
 \ifx #1\expandafter \@firstoftwo
 \else \expandafter \@secondoftwo
 \fi
}%
\providecommand \natexlab [1]{#1}%
\providecommand \enquote  [1]{``#1''}%
\providecommand \bibnamefont  [1]{#1}%
\providecommand \bibfnamefont [1]{#1}%
\providecommand \citenamefont [1]{#1}%
\providecommand \href@noop [0]{\@secondoftwo}%
\providecommand \href [0]{\begingroup \@sanitize@url \@href}%
\providecommand \@href[1]{\@@startlink{#1}\@@href}%
\providecommand \@@href[1]{\endgroup#1\@@endlink}%
\providecommand \@sanitize@url [0]{\catcode `\\12\catcode `\$12\catcode
  `\&12\catcode `\#12\catcode `\^12\catcode `\_12\catcode `\%12\relax}%
\providecommand \@@startlink[1]{}%
\providecommand \@@endlink[0]{}%
\providecommand \url  [0]{\begingroup\@sanitize@url \@url }%
\providecommand \@url [1]{\endgroup\@href {#1}{\urlprefix }}%
\providecommand \urlprefix  [0]{URL }%
\providecommand \Eprint [0]{\href }%
\providecommand \doibase [0]{http://dx.doi.org/}%
\providecommand \selectlanguage [0]{\@gobble}%
\providecommand \bibinfo  [0]{\@secondoftwo}%
\providecommand \bibfield  [0]{\@secondoftwo}%
\providecommand \translation [1]{[#1]}%
\providecommand \BibitemOpen [0]{}%
\providecommand \bibitemStop [0]{}%
\providecommand \bibitemNoStop [0]{.\EOS\space}%
\providecommand \EOS [0]{\spacefactor3000\relax}%
\providecommand \BibitemShut  [1]{\csname bibitem#1\endcsname}%
\let\auto@bib@innerbib\@empty
\bibitem [{\citenamefont {Hasan}\ and\ \citenamefont {Kane}(2010)}]{Hasan2010}%
  \BibitemOpen
  \bibfield  {author} {\bibinfo {author} {\bibfnamefont {M.~Z.}\ \bibnamefont
  {Hasan}}\ and\ \bibinfo {author} {\bibfnamefont {C.~L.}\ \bibnamefont
  {Kane}},\ }\href {\doibase 10.1103/RevModPhys.82.3045} {\bibfield  {journal}
  {\bibinfo  {journal} {Rev. Mod. Phys.}\ }\textbf {\bibinfo {volume} {82}},\
  \bibinfo {pages} {3045} (\bibinfo {year} {2010})}\BibitemShut {NoStop}%
\bibitem [{\citenamefont {Qi}\ and\ \citenamefont {Zhang}(2011)}]{Qi2011}%
  \BibitemOpen
  \bibfield  {author} {\bibinfo {author} {\bibfnamefont {X.-L.}\ \bibnamefont
  {Qi}}\ and\ \bibinfo {author} {\bibfnamefont {S.-C.}\ \bibnamefont {Zhang}},\
  }\href {\doibase 10.1103/RevModPhys.83.1057} {\bibfield  {journal} {\bibinfo
  {journal} {Rev. Mod. Phys.}\ }\textbf {\bibinfo {volume} {83}},\ \bibinfo
  {pages} {1057} (\bibinfo {year} {2011})}\BibitemShut {NoStop}%
\bibitem [{\citenamefont {Klitzing}\ \emph {et~al.}(1980)\citenamefont
  {Klitzing}, \citenamefont {Dorda},\ and\ \citenamefont
  {Pepper}}]{Klitzing1980}%
  \BibitemOpen
  \bibfield  {author} {\bibinfo {author} {\bibfnamefont {K.~v.}\ \bibnamefont
  {Klitzing}}, \bibinfo {author} {\bibfnamefont {G.}~\bibnamefont {Dorda}}, \
  and\ \bibinfo {author} {\bibfnamefont {M.}~\bibnamefont {Pepper}},\ }\href
  {\doibase 10.1103/PhysRevLett.45.494} {\bibfield  {journal} {\bibinfo
  {journal} {Phys. Rev. Lett.}\ }\textbf {\bibinfo {volume} {45}},\ \bibinfo
  {pages} {494} (\bibinfo {year} {1980})}\BibitemShut {NoStop}%
\bibitem [{\citenamefont {Haldane}(1988)}]{Haldane1988}%
  \BibitemOpen
  \bibfield  {author} {\bibinfo {author} {\bibfnamefont {F.~D.~M.}\
  \bibnamefont {Haldane}},\ }\href {\doibase 10.1103/PhysRevLett.61.2015}
  {\bibfield  {journal} {\bibinfo  {journal} {Phys. Rev. Lett.}\ }\textbf
  {\bibinfo {volume} {61}},\ \bibinfo {pages} {2015} (\bibinfo {year}
  {1988})}\BibitemShut {NoStop}%
\bibitem [{\citenamefont {Thouless}\ \emph {et~al.}(1982)\citenamefont
  {Thouless}, \citenamefont {Kohmoto}, \citenamefont {Nightingale},\ and\
  \citenamefont {den Nijs}}]{Thouless1982}%
  \BibitemOpen
  \bibfield  {author} {\bibinfo {author} {\bibfnamefont {D.~J.}\ \bibnamefont
  {Thouless}}, \bibinfo {author} {\bibfnamefont {M.}~\bibnamefont {Kohmoto}},
  \bibinfo {author} {\bibfnamefont {M.~P.}\ \bibnamefont {Nightingale}}, \ and\
  \bibinfo {author} {\bibfnamefont {M.}~\bibnamefont {den Nijs}},\ }\href
  {\doibase 10.1103/PhysRevLett.49.405} {\bibfield  {journal} {\bibinfo
  {journal} {Phys. Rev. Lett.}\ }\textbf {\bibinfo {volume} {49}},\ \bibinfo
  {pages} {405} (\bibinfo {year} {1982})}\BibitemShut {NoStop}%
\bibitem [{\citenamefont {Avron}\ \emph {et~al.}(1983)\citenamefont {Avron},
  \citenamefont {Seiler},\ and\ \citenamefont {Simon}}]{Avron1983}%
  \BibitemOpen
  \bibfield  {author} {\bibinfo {author} {\bibfnamefont {J.~E.}\ \bibnamefont
  {Avron}}, \bibinfo {author} {\bibfnamefont {R.}~\bibnamefont {Seiler}}, \
  and\ \bibinfo {author} {\bibfnamefont {B.}~\bibnamefont {Simon}},\ }\href
  {\doibase 10.1103/PhysRevLett.51.51} {\bibfield  {journal} {\bibinfo
  {journal} {Phys. Rev. Lett.}\ }\textbf {\bibinfo {volume} {51}},\ \bibinfo
  {pages} {51} (\bibinfo {year} {1983})}\BibitemShut {NoStop}%
\bibitem [{\citenamefont {Kane}\ and\ \citenamefont
  {Mele}(2005{\natexlab{a}})}]{Kane2005a}%
  \BibitemOpen
  \bibfield  {author} {\bibinfo {author} {\bibfnamefont {C.~L.}\ \bibnamefont
  {Kane}}\ and\ \bibinfo {author} {\bibfnamefont {E.~J.}\ \bibnamefont
  {Mele}},\ }\href {\doibase 10.1103/PhysRevLett.95.226801} {\bibfield
  {journal} {\bibinfo  {journal} {Phys. Rev. Lett.}\ }\textbf {\bibinfo
  {volume} {95}},\ \bibinfo {pages} {226801} (\bibinfo {year}
  {2005}{\natexlab{a}})}\BibitemShut {NoStop}%
\bibitem [{\citenamefont {Kane}\ and\ \citenamefont
  {Mele}(2005{\natexlab{b}})}]{Kane2005b}%
  \BibitemOpen
  \bibfield  {author} {\bibinfo {author} {\bibfnamefont {C.~L.}\ \bibnamefont
  {Kane}}\ and\ \bibinfo {author} {\bibfnamefont {E.~J.}\ \bibnamefont
  {Mele}},\ }\href {\doibase 10.1103/PhysRevLett.95.146802} {\bibfield
  {journal} {\bibinfo  {journal} {Phys. Rev. Lett.}\ }\textbf {\bibinfo
  {volume} {95}},\ \bibinfo {pages} {146802} (\bibinfo {year}
  {2005}{\natexlab{b}})}\BibitemShut {NoStop}%
\bibitem [{\citenamefont {Fu}\ and\ \citenamefont {Kane}(2006)}]{Fu2006time}%
  \BibitemOpen
  \bibfield  {author} {\bibinfo {author} {\bibfnamefont {L.}~\bibnamefont
  {Fu}}\ and\ \bibinfo {author} {\bibfnamefont {C.~L.}\ \bibnamefont {Kane}},\
  }\href {\doibase 10.1103/PhysRevB.74.195312} {\bibfield  {journal} {\bibinfo
  {journal} {Phys. Rev. B}\ }\textbf {\bibinfo {volume} {74}},\ \bibinfo
  {pages} {195312} (\bibinfo {year} {2006})}\BibitemShut {NoStop}%
\bibitem [{\citenamefont {Sheng}\ \emph {et~al.}(2006)\citenamefont {Sheng},
  \citenamefont {Weng}, \citenamefont {Sheng},\ and\ \citenamefont
  {Haldane}}]{Sheng2006spin}%
  \BibitemOpen
  \bibfield  {author} {\bibinfo {author} {\bibfnamefont {D.~N.}\ \bibnamefont
  {Sheng}}, \bibinfo {author} {\bibfnamefont {Z.~Y.}\ \bibnamefont {Weng}},
  \bibinfo {author} {\bibfnamefont {L.}~\bibnamefont {Sheng}}, \ and\ \bibinfo
  {author} {\bibfnamefont {F.~D.~M.}\ \bibnamefont {Haldane}},\ }\href
  {\doibase 10.1103/PhysRevLett.97.036808} {\bibfield  {journal} {\bibinfo
  {journal} {Phys. Rev. Lett.}\ }\textbf {\bibinfo {volume} {97}},\ \bibinfo
  {pages} {036808} (\bibinfo {year} {2006})}\BibitemShut {NoStop}%
\bibitem [{\citenamefont {Prodan}(2009)}]{Prodan2009}%
  \BibitemOpen
  \bibfield  {author} {\bibinfo {author} {\bibfnamefont {E.}~\bibnamefont
  {Prodan}},\ }\href {\doibase 10.1103/PhysRevB.80.125327} {\bibfield
  {journal} {\bibinfo  {journal} {Phys. Rev. B}\ }\textbf {\bibinfo {volume}
  {80}},\ \bibinfo {pages} {125327} (\bibinfo {year} {2009})}\BibitemShut
  {NoStop}%
\bibitem [{\citenamefont {Fu}\ and\ \citenamefont {Kane}(2007)}]{Fu2007b}%
  \BibitemOpen
  \bibfield  {author} {\bibinfo {author} {\bibfnamefont {L.}~\bibnamefont
  {Fu}}\ and\ \bibinfo {author} {\bibfnamefont {C.~L.}\ \bibnamefont {Kane}},\
  }\href {\doibase 10.1103/PhysRevB.76.045302} {\bibfield  {journal} {\bibinfo
  {journal} {Phys. Rev. B}\ }\textbf {\bibinfo {volume} {76}},\ \bibinfo
  {pages} {045302} (\bibinfo {year} {2007})}\BibitemShut {NoStop}%
\bibitem [{\citenamefont {So}(1985)}]{So1985}%
  \BibitemOpen
  \bibfield  {author} {\bibinfo {author} {\bibfnamefont {H.}~\bibnamefont
  {So}},\ }\href {\doibase 10.1143/PTP.74.585} {\bibfield  {journal} {\bibinfo
  {journal} {Progress of Theoretical Physics}\ }\textbf {\bibinfo {volume}
  {74}},\ \bibinfo {pages} {585} (\bibinfo {year} {1985})}\BibitemShut
  {NoStop}%
\bibitem [{\citenamefont {Ishikawa}\ and\ \citenamefont
  {Matsuyama}(1986)}]{Ishikawa}%
  \BibitemOpen
  \bibfield  {author} {\bibinfo {author} {\bibfnamefont {K.}~\bibnamefont
  {Ishikawa}}\ and\ \bibinfo {author} {\bibfnamefont {T.}~\bibnamefont
  {Matsuyama}},\ }\href {\doibase 10.1007/BF01410451} {\bibfield  {journal}
  {\bibinfo  {journal} {Z. Phys. C}\ }\textbf {\bibinfo {volume} {33}},\
  \bibinfo {pages} {41} (\bibinfo {year} {1986})}\BibitemShut {NoStop}%
\bibitem [{\citenamefont {Volovik}(1988)}]{Volovik1988}%
  \BibitemOpen
  \bibfield  {author} {\bibinfo {author} {\bibfnamefont {G.~E.}\ \bibnamefont
  {Volovik}},\ }\href {http://www.jetp.ac.ru/cgi-bin/dn/e_067_09_1804.pdf}
  {\bibfield  {journal} {\bibinfo  {journal} {JETP}\ }\textbf {\bibinfo
  {volume} {67}},\ \bibinfo {pages} {1804} (\bibinfo {year}
  {1988})}\BibitemShut {NoStop}%
\bibitem [{\citenamefont {Wang}\ \emph {et~al.}(2010)\citenamefont {Wang},
  \citenamefont {Qi},\ and\ \citenamefont {Zhang}}]{Wang2010}%
  \BibitemOpen
  \bibfield  {author} {\bibinfo {author} {\bibfnamefont {Z.}~\bibnamefont
  {Wang}}, \bibinfo {author} {\bibfnamefont {X.-L.}\ \bibnamefont {Qi}}, \ and\
  \bibinfo {author} {\bibfnamefont {S.-C.}\ \bibnamefont {Zhang}},\ }\href
  {\doibase 10.1103/PhysRevLett.105.256803} {\bibfield  {journal} {\bibinfo
  {journal} {Phys. Rev. Lett.}\ }\textbf {\bibinfo {volume} {105}},\ \bibinfo
  {pages} {256803} (\bibinfo {year} {2010})}\BibitemShut {NoStop}%
\bibitem [{\citenamefont {Gurarie}(2011)}]{Gurarie2011}%
  \BibitemOpen
  \bibfield  {author} {\bibinfo {author} {\bibfnamefont {V.}~\bibnamefont
  {Gurarie}},\ }\href {\doibase 10.1103/PhysRevB.83.085426} {\bibfield
  {journal} {\bibinfo  {journal} {Phys. Rev. B}\ }\textbf {\bibinfo {volume}
  {83}},\ \bibinfo {pages} {085426} (\bibinfo {year} {2011})}\BibitemShut
  {NoStop}%
\bibitem [{\citenamefont {Volovik}(2009)}]{volovik2009universe}%
  \BibitemOpen
  \bibfield  {author} {\bibinfo {author} {\bibfnamefont {G.~E.}\ \bibnamefont
  {Volovik}},\ }\href@noop {} {\emph {\bibinfo {title} {The universe in a
  helium droplet}}}\ (\bibinfo  {publisher} {Oxford University Press New
  York},\ \bibinfo {year} {2009})\BibitemShut {NoStop}%
\bibitem [{\citenamefont {Niu}\ \emph {et~al.}(1985)\citenamefont {Niu},
  \citenamefont {Thouless},\ and\ \citenamefont {Wu}}]{Niu1985}%
  \BibitemOpen
  \bibfield  {author} {\bibinfo {author} {\bibfnamefont {Q.}~\bibnamefont
  {Niu}}, \bibinfo {author} {\bibfnamefont {D.~J.}\ \bibnamefont {Thouless}}, \
  and\ \bibinfo {author} {\bibfnamefont {Y.-S.}\ \bibnamefont {Wu}},\ }\href
  {\doibase 10.1103/PhysRevB.31.3372} {\bibfield  {journal} {\bibinfo
  {journal} {Phys. Rev. B}\ }\textbf {\bibinfo {volume} {31}},\ \bibinfo
  {pages} {3372} (\bibinfo {year} {1985})}\BibitemShut {NoStop}%
\bibitem [{\citenamefont {Wang}\ and\ \citenamefont {Zhang}(2014)}]{Wang2014}%
  \BibitemOpen
  \bibfield  {author} {\bibinfo {author} {\bibfnamefont {Z.}~\bibnamefont
  {Wang}}\ and\ \bibinfo {author} {\bibfnamefont {S.-C.}\ \bibnamefont
  {Zhang}},\ }\href {\doibase 10.1103/PhysRevX.4.011006} {\bibfield  {journal}
  {\bibinfo  {journal} {Phys. Rev. X}\ }\textbf {\bibinfo {volume} {4}},\
  \bibinfo {pages} {011006} (\bibinfo {year} {2014})}\BibitemShut {NoStop}%
\bibitem [{\citenamefont {Wang}\ and\ \citenamefont
  {Zhang}(2012{\natexlab{a}})}]{Wang2012c}%
  \BibitemOpen
  \bibfield  {author} {\bibinfo {author} {\bibfnamefont {Z.}~\bibnamefont
  {Wang}}\ and\ \bibinfo {author} {\bibfnamefont {S.-C.}\ \bibnamefont
  {Zhang}},\ }\href {\doibase 10.1103/PhysRevX.2.031008} {\bibfield  {journal}
  {\bibinfo  {journal} {Phys. Rev. X}\ }\textbf {\bibinfo {volume} {2}},\
  \bibinfo {pages} {031008} (\bibinfo {year} {2012}{\natexlab{a}})}\BibitemShut
  {NoStop}%
\bibitem [{\citenamefont {{Wang}}\ and\ \citenamefont
  {{Yan}}(2013)}]{Wang2013topological}%
  \BibitemOpen
  \bibfield  {author} {\bibinfo {author} {\bibfnamefont {Z.}~\bibnamefont
  {{Wang}}}\ and\ \bibinfo {author} {\bibfnamefont {B.}~\bibnamefont {{Yan}}},\
  }\href {\doibase 10.1088/0953-8984/25/15/155601} {\bibfield  {journal}
  {\bibinfo  {journal} {Journal of Physics: Condensed Matter}\ }\textbf
  {\bibinfo {volume} {25}},\ \bibinfo {eid} {155601} (\bibinfo {year}
  {2013})}\BibitemShut {NoStop}%
\bibitem [{\citenamefont {Yoshida}\ \emph {et~al.}(2014)\citenamefont
  {Yoshida}, \citenamefont {Peters}, \citenamefont {Fujimoto},\ and\
  \citenamefont {Kawakami}}]{Yoshida2014}%
  \BibitemOpen
  \bibfield  {author} {\bibinfo {author} {\bibfnamefont {T.}~\bibnamefont
  {Yoshida}}, \bibinfo {author} {\bibfnamefont {R.}~\bibnamefont {Peters}},
  \bibinfo {author} {\bibfnamefont {S.}~\bibnamefont {Fujimoto}}, \ and\
  \bibinfo {author} {\bibfnamefont {N.}~\bibnamefont {Kawakami}},\ }\href
  {\doibase 10.1103/PhysRevLett.112.196404} {\bibfield  {journal} {\bibinfo
  {journal} {Phys. Rev. Lett.}\ }\textbf {\bibinfo {volume} {112}},\ \bibinfo
  {pages} {196404} (\bibinfo {year} {2014})}\BibitemShut {NoStop}%
\bibitem [{\citenamefont {Lu}\ \emph {et~al.}(2013)\citenamefont {Lu},
  \citenamefont {Zhao}, \citenamefont {Weng}, \citenamefont {Fang},\ and\
  \citenamefont {Dai}}]{Lu2013}%
  \BibitemOpen
  \bibfield  {author} {\bibinfo {author} {\bibfnamefont {F.}~\bibnamefont
  {Lu}}, \bibinfo {author} {\bibfnamefont {J.}~\bibnamefont {Zhao}}, \bibinfo
  {author} {\bibfnamefont {H.}~\bibnamefont {Weng}}, \bibinfo {author}
  {\bibfnamefont {Z.}~\bibnamefont {Fang}}, \ and\ \bibinfo {author}
  {\bibfnamefont {X.}~\bibnamefont {Dai}},\ }\href {\doibase
  10.1103/PhysRevLett.110.096401} {\bibfield  {journal} {\bibinfo  {journal}
  {Phys. Rev. Lett.}\ }\textbf {\bibinfo {volume} {110}},\ \bibinfo {pages}
  {096401} (\bibinfo {year} {2013})}\BibitemShut {NoStop}%
\bibitem [{\citenamefont {Deng}\ \emph {et~al.}(2013)\citenamefont {Deng},
  \citenamefont {Haule},\ and\ \citenamefont {Kotliar}}]{Deng2013Plutonium}%
  \BibitemOpen
  \bibfield  {author} {\bibinfo {author} {\bibfnamefont {X.}~\bibnamefont
  {Deng}}, \bibinfo {author} {\bibfnamefont {K.}~\bibnamefont {Haule}}, \ and\
  \bibinfo {author} {\bibfnamefont {G.}~\bibnamefont {Kotliar}},\ }\href
  {\doibase 10.1103/PhysRevLett.111.176404} {\bibfield  {journal} {\bibinfo
  {journal} {Phys. Rev. Lett.}\ }\textbf {\bibinfo {volume} {111}},\ \bibinfo
  {pages} {176404} (\bibinfo {year} {2013})}\BibitemShut {NoStop}%
\bibitem [{\citenamefont {Hung}\ \emph {et~al.}(2013)\citenamefont {Hung},
  \citenamefont {Wang}, \citenamefont {Gu},\ and\ \citenamefont
  {Fiete}}]{Hung2013}%
  \BibitemOpen
  \bibfield  {author} {\bibinfo {author} {\bibfnamefont {H.-H.}\ \bibnamefont
  {Hung}}, \bibinfo {author} {\bibfnamefont {L.}~\bibnamefont {Wang}}, \bibinfo
  {author} {\bibfnamefont {Z.-C.}\ \bibnamefont {Gu}}, \ and\ \bibinfo {author}
  {\bibfnamefont {G.~A.}\ \bibnamefont {Fiete}},\ }\href {\doibase
  10.1103/PhysRevB.87.121113} {\bibfield  {journal} {\bibinfo  {journal} {Phys.
  Rev. B}\ }\textbf {\bibinfo {volume} {87}},\ \bibinfo {pages} {121113}
  (\bibinfo {year} {2013})}\BibitemShut {NoStop}%
\bibitem [{\citenamefont {Lang}\ \emph {et~al.}(2013)\citenamefont {Lang},
  \citenamefont {Essin}, \citenamefont {Gurarie},\ and\ \citenamefont
  {Wessel}}]{Lang2013}%
  \BibitemOpen
  \bibfield  {author} {\bibinfo {author} {\bibfnamefont {T.~C.}\ \bibnamefont
  {Lang}}, \bibinfo {author} {\bibfnamefont {A.~M.}\ \bibnamefont {Essin}},
  \bibinfo {author} {\bibfnamefont {V.}~\bibnamefont {Gurarie}}, \ and\
  \bibinfo {author} {\bibfnamefont {S.}~\bibnamefont {Wessel}},\ }\href
  {\doibase 10.1103/PhysRevB.87.205101} {\bibfield  {journal} {\bibinfo
  {journal} {Phys. Rev. B}\ }\textbf {\bibinfo {volume} {87}},\ \bibinfo
  {pages} {205101} (\bibinfo {year} {2013})}\BibitemShut {NoStop}%
\bibitem [{\citenamefont {Hung}\ \emph {et~al.}(2014)\citenamefont {Hung},
  \citenamefont {Chua}, \citenamefont {Wang},\ and\ \citenamefont
  {Fiete}}]{Hung2014}%
  \BibitemOpen
  \bibfield  {author} {\bibinfo {author} {\bibfnamefont {H.-H.}\ \bibnamefont
  {Hung}}, \bibinfo {author} {\bibfnamefont {V.}~\bibnamefont {Chua}}, \bibinfo
  {author} {\bibfnamefont {L.}~\bibnamefont {Wang}}, \ and\ \bibinfo {author}
  {\bibfnamefont {G.~A.}\ \bibnamefont {Fiete}},\ }\href {\doibase
  10.1103/PhysRevB.89.235104} {\bibfield  {journal} {\bibinfo  {journal} {Phys.
  Rev. B}\ }\textbf {\bibinfo {volume} {89}},\ \bibinfo {pages} {235104}
  (\bibinfo {year} {2014})}\BibitemShut {NoStop}%
\bibitem [{\citenamefont {Meng}\ \emph {et~al.}(2014)\citenamefont {Meng},
  \citenamefont {Hung},\ and\ \citenamefont {Lang}}]{Meng2014}%
  \BibitemOpen
  \bibfield  {author} {\bibinfo {author} {\bibfnamefont {Z.~Y.}\ \bibnamefont
  {Meng}}, \bibinfo {author} {\bibfnamefont {H.-H.}\ \bibnamefont {Hung}}, \
  and\ \bibinfo {author} {\bibfnamefont {T.~C.}\ \bibnamefont {Lang}},\ }\href
  {\doibase 10.1142/S0217984914300014} {\bibfield  {journal} {\bibinfo
  {journal} {Modern Physics Letters B}\ }\textbf {\bibinfo {volume} {28}},\
  \bibinfo {pages} {1430001} (\bibinfo {year} {2014})}\BibitemShut {NoStop}%
\bibitem [{\citenamefont {Grandi}\ \emph
  {et~al.}(2015{\natexlab{a}})\citenamefont {Grandi}, \citenamefont {Manghi},
  \citenamefont {Corradini},\ and\ \citenamefont {Bertoni}}]{Grandi2015a}%
  \BibitemOpen
  \bibfield  {author} {\bibinfo {author} {\bibfnamefont {F.}~\bibnamefont
  {Grandi}}, \bibinfo {author} {\bibfnamefont {F.}~\bibnamefont {Manghi}},
  \bibinfo {author} {\bibfnamefont {O.}~\bibnamefont {Corradini}}, \ and\
  \bibinfo {author} {\bibfnamefont {C.~M.}\ \bibnamefont {Bertoni}},\ }\href
  {\doibase 10.1103/PhysRevB.91.115112} {\bibfield  {journal} {\bibinfo
  {journal} {Phys. Rev. B}\ }\textbf {\bibinfo {volume} {91}},\ \bibinfo
  {pages} {115112} (\bibinfo {year} {2015}{\natexlab{a}})}\BibitemShut
  {NoStop}%
\bibitem [{\citenamefont {Grandi}\ \emph
  {et~al.}(2015{\natexlab{b}})\citenamefont {Grandi}, \citenamefont {Manghi},
  \citenamefont {Corradini}, \citenamefont {Bertoni},\ and\ \citenamefont
  {Bonini}}]{Grandi2015b}%
  \BibitemOpen
  \bibfield  {author} {\bibinfo {author} {\bibfnamefont {F.}~\bibnamefont
  {Grandi}}, \bibinfo {author} {\bibfnamefont {F.}~\bibnamefont {Manghi}},
  \bibinfo {author} {\bibfnamefont {O.}~\bibnamefont {Corradini}}, \bibinfo
  {author} {\bibfnamefont {C.~M.}\ \bibnamefont {Bertoni}}, \ and\ \bibinfo
  {author} {\bibfnamefont {A.}~\bibnamefont {Bonini}},\ }\href
  {http://stacks.iop.org/1367-2630/17/i=2/a=023004} {\bibfield  {journal}
  {\bibinfo  {journal} {New Journal of Physics}\ }\textbf {\bibinfo {volume}
  {17}},\ \bibinfo {pages} {023004} (\bibinfo {year}
  {2015}{\natexlab{b}})}\BibitemShut {NoStop}%
\bibitem [{\citenamefont {Budich}\ \emph {et~al.}(2012)\citenamefont {Budich},
  \citenamefont {Thomale}, \citenamefont {Li}, \citenamefont {Laubach},\ and\
  \citenamefont {Zhang}}]{Budich2012fluctuation}%
  \BibitemOpen
  \bibfield  {author} {\bibinfo {author} {\bibfnamefont {J.~C.}\ \bibnamefont
  {Budich}}, \bibinfo {author} {\bibfnamefont {R.}~\bibnamefont {Thomale}},
  \bibinfo {author} {\bibfnamefont {G.}~\bibnamefont {Li}}, \bibinfo {author}
  {\bibfnamefont {M.}~\bibnamefont {Laubach}}, \ and\ \bibinfo {author}
  {\bibfnamefont {S.-C.}\ \bibnamefont {Zhang}},\ }\href {\doibase
  10.1103/PhysRevB.86.201407} {\bibfield  {journal} {\bibinfo  {journal} {Phys.
  Rev. B}\ }\textbf {\bibinfo {volume} {86}},\ \bibinfo {pages} {201407}
  (\bibinfo {year} {2012})}\BibitemShut {NoStop}%
\bibitem [{\citenamefont {Budich}\ \emph {et~al.}(2013)\citenamefont {Budich},
  \citenamefont {Trauzettel},\ and\ \citenamefont {Sangiovanni}}]{Budich2013}%
  \BibitemOpen
  \bibfield  {author} {\bibinfo {author} {\bibfnamefont {J.~C.}\ \bibnamefont
  {Budich}}, \bibinfo {author} {\bibfnamefont {B.}~\bibnamefont {Trauzettel}},
  \ and\ \bibinfo {author} {\bibfnamefont {G.}~\bibnamefont {Sangiovanni}},\
  }\href {\doibase 10.1103/PhysRevB.87.235104} {\bibfield  {journal} {\bibinfo
  {journal} {Phys. Rev. B}\ }\textbf {\bibinfo {volume} {87}},\ \bibinfo
  {pages} {235104} (\bibinfo {year} {2013})}\BibitemShut {NoStop}%
\bibitem [{\citenamefont {Amaricci}\ \emph {et~al.}(2015)\citenamefont
  {Amaricci}, \citenamefont {Budich}, \citenamefont {Capone}, \citenamefont
  {Trauzettel},\ and\ \citenamefont {Sangiovanni}}]{Amaricci2015}%
  \BibitemOpen
  \bibfield  {author} {\bibinfo {author} {\bibfnamefont {A.}~\bibnamefont
  {Amaricci}}, \bibinfo {author} {\bibfnamefont {J.~C.}\ \bibnamefont
  {Budich}}, \bibinfo {author} {\bibfnamefont {M.}~\bibnamefont {Capone}},
  \bibinfo {author} {\bibfnamefont {B.}~\bibnamefont {Trauzettel}}, \ and\
  \bibinfo {author} {\bibfnamefont {G.}~\bibnamefont {Sangiovanni}},\ }\href
  {\doibase 10.1103/PhysRevLett.114.185701} {\bibfield  {journal} {\bibinfo
  {journal} {Phys. Rev. Lett.}\ }\textbf {\bibinfo {volume} {114}},\ \bibinfo
  {pages} {185701} (\bibinfo {year} {2015})}\BibitemShut {NoStop}%
\bibitem [{\citenamefont {Chen}\ \emph {et~al.}(2015)\citenamefont {Chen},
  \citenamefont {Hung}, \citenamefont {Su}, \citenamefont {Fiete},\ and\
  \citenamefont {Ting}}]{Chen2015}%
  \BibitemOpen
  \bibfield  {author} {\bibinfo {author} {\bibfnamefont {Y.-H.}\ \bibnamefont
  {Chen}}, \bibinfo {author} {\bibfnamefont {H.-H.}\ \bibnamefont {Hung}},
  \bibinfo {author} {\bibfnamefont {G.}~\bibnamefont {Su}}, \bibinfo {author}
  {\bibfnamefont {G.~A.}\ \bibnamefont {Fiete}}, \ and\ \bibinfo {author}
  {\bibfnamefont {C.~S.}\ \bibnamefont {Ting}},\ }\href {\doibase
  10.1103/PhysRevB.91.045122} {\bibfield  {journal} {\bibinfo  {journal} {Phys.
  Rev. B}\ }\textbf {\bibinfo {volume} {91}},\ \bibinfo {pages} {045122}
  (\bibinfo {year} {2015})}\BibitemShut {NoStop}%
\bibitem [{\citenamefont {Maier}\ \emph {et~al.}(2005)\citenamefont {Maier},
  \citenamefont {Jarrell}, \citenamefont {Pruschke},\ and\ \citenamefont
  {Hettler}}]{Maier2005}%
  \BibitemOpen
  \bibfield  {author} {\bibinfo {author} {\bibfnamefont {T.}~\bibnamefont
  {Maier}}, \bibinfo {author} {\bibfnamefont {M.}~\bibnamefont {Jarrell}},
  \bibinfo {author} {\bibfnamefont {T.}~\bibnamefont {Pruschke}}, \ and\
  \bibinfo {author} {\bibfnamefont {M.~H.}\ \bibnamefont {Hettler}},\ }\href
  {\doibase 10.1103/RevModPhys.77.1027} {\bibfield  {journal} {\bibinfo
  {journal} {Rev. Mod. Phys.}\ }\textbf {\bibinfo {volume} {77}},\ \bibinfo
  {pages} {1027} (\bibinfo {year} {2005})}\BibitemShut {NoStop}%
\bibitem [{\citenamefont {Wang}\ \emph {et~al.}(2011)\citenamefont {Wang},
  \citenamefont {Qi},\ and\ \citenamefont {Zhang}}]{Wang2011}%
  \BibitemOpen
  \bibfield  {author} {\bibinfo {author} {\bibfnamefont {Z.}~\bibnamefont
  {Wang}}, \bibinfo {author} {\bibfnamefont {X.-L.}\ \bibnamefont {Qi}}, \ and\
  \bibinfo {author} {\bibfnamefont {S.-C.}\ \bibnamefont {Zhang}},\ }\href
  {\doibase 10.1103/PhysRevB.84.014527} {\bibfield  {journal} {\bibinfo
  {journal} {Phys. Rev. B}\ }\textbf {\bibinfo {volume} {84}},\ \bibinfo
  {pages} {014527} (\bibinfo {year} {2011})}\BibitemShut {NoStop}%
\bibitem [{\citenamefont {Wang}\ \emph {et~al.}(2012)\citenamefont {Wang},
  \citenamefont {Qi},\ and\ \citenamefont {Zhang}}]{Wang2012a}%
  \BibitemOpen
  \bibfield  {author} {\bibinfo {author} {\bibfnamefont {Z.}~\bibnamefont
  {Wang}}, \bibinfo {author} {\bibfnamefont {X.-L.}\ \bibnamefont {Qi}}, \ and\
  \bibinfo {author} {\bibfnamefont {S.-C.}\ \bibnamefont {Zhang}},\ }\href
  {\doibase 10.1103/PhysRevB.85.165126} {\bibfield  {journal} {\bibinfo
  {journal} {Phys. Rev. B}\ }\textbf {\bibinfo {volume} {85}},\ \bibinfo
  {pages} {165126} (\bibinfo {year} {2012})}\BibitemShut {NoStop}%
\bibitem [{\citenamefont {Wang}\ and\ \citenamefont
  {Zhang}(2012{\natexlab{b}})}]{Wang2012b}%
  \BibitemOpen
  \bibfield  {author} {\bibinfo {author} {\bibfnamefont {Z.}~\bibnamefont
  {Wang}}\ and\ \bibinfo {author} {\bibfnamefont {S.-C.}\ \bibnamefont
  {Zhang}},\ }\href {\doibase 10.1103/PhysRevB.86.165116} {\bibfield  {journal}
  {\bibinfo  {journal} {Phys. Rev. B}\ }\textbf {\bibinfo {volume} {86}},\
  \bibinfo {pages} {165116} (\bibinfo {year} {2012}{\natexlab{b}})}\BibitemShut
  {NoStop}%
\bibitem [{\citenamefont {Assaad}\ and\ \citenamefont
  {Evertz}(2008)}]{AssaadEvertz2008}%
  \BibitemOpen
  \bibfield  {author} {\bibinfo {author} {\bibfnamefont {F.}~\bibnamefont
  {Assaad}}\ and\ \bibinfo {author} {\bibfnamefont {H.}~\bibnamefont
  {Evertz}},\ }in\ \href {\doibase 10.1007/978-3-540-74686-7_10} {\emph
  {\bibinfo {booktitle} {Computational Many-Particle Physics}}},\ \bibinfo
  {series} {Lecture Notes in Physics}, Vol.\ \bibinfo {volume} {739},\ \bibinfo
  {editor} {edited by\ \bibinfo {editor} {\bibfnamefont {H.}~\bibnamefont
  {Fehske}}, \bibinfo {editor} {\bibfnamefont {R.}~\bibnamefont {Schneider}}, \
  and\ \bibinfo {editor} {\bibfnamefont {A.}~\bibnamefont {Wei{\ss}e}}}\
  (\bibinfo  {publisher} {Springer Berlin Heidelberg},\ \bibinfo {year}
  {2008})\ pp.\ \bibinfo {pages} {277--356}\BibitemShut {NoStop}%
\bibitem [{\citenamefont {Parcollet}\ \emph {et~al.}(2004)\citenamefont
  {Parcollet}, \citenamefont {Biroli},\ and\ \citenamefont
  {Kotliar}}]{Parcollet2004}%
  \BibitemOpen
  \bibfield  {author} {\bibinfo {author} {\bibfnamefont {O.}~\bibnamefont
  {Parcollet}}, \bibinfo {author} {\bibfnamefont {G.}~\bibnamefont {Biroli}}, \
  and\ \bibinfo {author} {\bibfnamefont {G.}~\bibnamefont {Kotliar}},\ }\href
  {\doibase 10.1103/PhysRevLett.92.226402} {\bibfield  {journal} {\bibinfo
  {journal} {Phys. Rev. Lett.}\ }\textbf {\bibinfo {volume} {92}},\ \bibinfo
  {pages} {226402} (\bibinfo {year} {2004})}\BibitemShut {NoStop}%
\bibitem [{\citenamefont {Biroli}\ \emph {et~al.}(2004)\citenamefont {Biroli},
  \citenamefont {Parcollet},\ and\ \citenamefont {Kotliar}}]{Biroli2004}%
  \BibitemOpen
  \bibfield  {author} {\bibinfo {author} {\bibfnamefont {G.}~\bibnamefont
  {Biroli}}, \bibinfo {author} {\bibfnamefont {O.}~\bibnamefont {Parcollet}}, \
  and\ \bibinfo {author} {\bibfnamefont {G.}~\bibnamefont {Kotliar}},\ }\href
  {\doibase 10.1103/PhysRevB.69.205108} {\bibfield  {journal} {\bibinfo
  {journal} {Phys. Rev. B}\ }\textbf {\bibinfo {volume} {69}},\ \bibinfo
  {pages} {205108} (\bibinfo {year} {2004})}\BibitemShut {NoStop}%
\bibitem [{\citenamefont {Sakai}\ \emph {et~al.}(2012)\citenamefont {Sakai},
  \citenamefont {Sangiovanni}, \citenamefont {Civelli}, \citenamefont {Motome},
  \citenamefont {Held},\ and\ \citenamefont {Imada}}]{Sakai2012}%
  \BibitemOpen
  \bibfield  {author} {\bibinfo {author} {\bibfnamefont {S.}~\bibnamefont
  {Sakai}}, \bibinfo {author} {\bibfnamefont {G.}~\bibnamefont {Sangiovanni}},
  \bibinfo {author} {\bibfnamefont {M.}~\bibnamefont {Civelli}}, \bibinfo
  {author} {\bibfnamefont {Y.}~\bibnamefont {Motome}}, \bibinfo {author}
  {\bibfnamefont {K.}~\bibnamefont {Held}}, \ and\ \bibinfo {author}
  {\bibfnamefont {M.}~\bibnamefont {Imada}},\ }\href {\doibase
  10.1103/PhysRevB.85.035102} {\bibfield  {journal} {\bibinfo  {journal} {Phys.
  Rev. B}\ }\textbf {\bibinfo {volume} {85}},\ \bibinfo {pages} {035102}
  (\bibinfo {year} {2012})}\BibitemShut {NoStop}%
\bibitem [{\citenamefont {Li}\ \emph {et~al.}(2015)\citenamefont {Li},
  \citenamefont {He},\ and\ \citenamefont {Lu}}]{QingXiao2015}%
  \BibitemOpen
  \bibfield  {author} {\bibinfo {author} {\bibfnamefont {Q.-X.}\ \bibnamefont
  {Li}}, \bibinfo {author} {\bibfnamefont {R.-Q.}\ \bibnamefont {He}}, \ and\
  \bibinfo {author} {\bibfnamefont {Z.-Y.}\ \bibnamefont {Lu}},\ }\href
  {\doibase 10.1103/PhysRevB.92.155127} {\bibfield  {journal} {\bibinfo
  {journal} {Phys. Rev. B}\ }\textbf {\bibinfo {volume} {92}},\ \bibinfo
  {pages} {155127} (\bibinfo {year} {2015})}\BibitemShut {NoStop}%
\bibitem [{\citenamefont {Wu}\ \emph {et~al.}(2015)\citenamefont {Wu},
  \citenamefont {He}, \citenamefont {You}, \citenamefont {Xu}, \citenamefont
  {Meng},\ and\ \citenamefont {Lu}}]{HQWu2015}%
  \BibitemOpen
  \bibfield  {author} {\bibinfo {author} {\bibfnamefont {H.-Q.}\ \bibnamefont
  {Wu}}, \bibinfo {author} {\bibfnamefont {Y.-Y.}\ \bibnamefont {He}}, \bibinfo
  {author} {\bibfnamefont {Y.-Z.}\ \bibnamefont {You}}, \bibinfo {author}
  {\bibfnamefont {C.}~\bibnamefont {Xu}}, \bibinfo {author} {\bibfnamefont
  {Z.~Y.}\ \bibnamefont {Meng}}, \ and\ \bibinfo {author} {\bibfnamefont
  {Z.-Y.}\ \bibnamefont {Lu}},\ }\href {http://arxiv.org/abs/1506.00549}
  {\bibfield  {journal} {\bibinfo  {journal} {arXiv}\ }\textbf {\bibinfo
  {volume} {1506}},\ \bibinfo {pages} {00549} (\bibinfo {year}
  {2015})}\BibitemShut {NoStop}%
\bibitem [{\citenamefont {Wu}\ \emph {et~al.}(2012)\citenamefont {Wu},
  \citenamefont {Rachel}, \citenamefont {Liu},\ and\ \citenamefont
  {Le~Hur}}]{Wu2012}%
  \BibitemOpen
  \bibfield  {author} {\bibinfo {author} {\bibfnamefont {W.}~\bibnamefont
  {Wu}}, \bibinfo {author} {\bibfnamefont {S.}~\bibnamefont {Rachel}}, \bibinfo
  {author} {\bibfnamefont {W.-M.}\ \bibnamefont {Liu}}, \ and\ \bibinfo
  {author} {\bibfnamefont {K.}~\bibnamefont {Le~Hur}},\ }\href {\doibase
  10.1103/PhysRevB.85.205102} {\bibfield  {journal} {\bibinfo  {journal} {Phys.
  Rev. B}\ }\textbf {\bibinfo {volume} {85}},\ \bibinfo {pages} {205102}
  (\bibinfo {year} {2012})}\BibitemShut {NoStop}%
\end{thebibliography}%

\end{document}